\documentclass[11pt,psfig]{article}
\usepackage{graphicx}
\usepackage{float}
\usepackage{subcaption}
\usepackage{amsmath,amssymb,mathtools,bbm}
\usepackage{bbold}
\usepackage{color}
\usepackage{cite}
\usepackage{tikz}
\usetikzlibrary{shapes}
\usepackage{longtable}
\usepackage{pgfplots}
\usepackage{subeqnarray}
\usepackage{dsfont}
\usepackage{changepage}
\tikzset{cross/.style={cross out, draw=black, minimum size=2*(#1-\pgflinewidth), inner sep=0pt, outer sep=0pt},
cross/.default={1pt}}

\title{Reconstruction of the phonon relaxation times using solutions of the Boltzmann transport equation }

\author{Mojtaba Forghani and Nicolas G. Hadjiconstantinou, \\Department of Mechanical Engineering, Massachusetts Institute of Technology\\ Cambridge, MA 02139, USA \\and\\ Jean-Philippe M. P\'{e}raud,\\Lawrence Berkeley Laboratory\\ Berkeley, CA 94720, USA}

\begin{document}
\maketitle 

\begin{abstract}
We present a method for reconstructing the phonon relaxation time distribution $\tau_{\omega}= \tau (\omega)$ (including polarization) in a material from thermal spectroscopy data. The distinguishing feature of this approach is that it does not make use of the effective thermal conductivity concept and associated approximations. The reconstruction is posed as an optimization problem in which the relaxation times $\tau_{\omega}= \tau (\omega)$ are determined by minimizing the discrepancy between the experimental relaxation traces and solutions of the Boltzmann transport equation (BTE) for the same problem. The latter may be analytical, in which case the procedure is very efficient, or numerical. The proposed method is illustrated using Monte Carlo solutions of thermal grating relaxation as synthetic experimental data. The reconstruction is shown to agree very well with the relaxation times used to generate the synthetic Monte Carlo data and remains robust in the presence of uncertainty (noise).
\end{abstract}

\section{Introduction}
\label{intro}
By probing the thermal response at timescales and lengthscales on the order of the phonon relaxation times and free paths, respectively, thermal spectroscopy provides a potentially powerful means of extracting information about these intrinsic transport properties of real materials \cite{ThermalSpectroscopy2011, NanoLetter2008, Cahill, Maznev2011, NatureCommun2014, McGaughey, Dames2015}. As a result, thermal spectroscopy has recently received a lot of attention for applications related to the development of thermoelectric materials \cite{ThermalSpectroscopy2011, Minnich2012, Lingping_nature}. 

The success of this technique relies on reliably extracting the information embedded in the material response, an inverse problem of considerable complexity that currently remains open. In order to use precise language, in this paper we focus on time-domain thermoreflectance. We believe that the methodology proposed in this paper can be straightforwardly extended to complementary approaches based on frequency-domain thermoreflectance \cite{Cahill, McGaughey, Dames2015}. 

Traditional approaches to time-domain thermoreflectance data analysis start by matching (in a least-squares-fit sense) the experimental response to numerical solutions of the heat conduction equation with the thermal conductivity treated as an adjustable, ``effective", quantity \cite{ThermalSpectroscopy2011}. The free path distribution is subsequently extracted by assuming that the fitted effective thermal conductivity is given by the convolution of the differential free path distribution in the material with a suppression function that is assumed to be geometry dependent \cite{Maznev2011, Vazrik2016}. An approximate suppression function has been developed for the thermal grating geometry \cite{Maznev2011, Hua2014}; suppression functions for  other, more complex, geometries have yet to be developed. Robust optimization frameworks for inverting the resulting integral equations have also been developed \cite{Minnich2012}.

However, since, by design, the material response in the experiment is not in the traditional Fourier regime---defined by  characteristic timescales and lengthscales being much longer than the phonon mean free time and mean free path, respectively---approaches based on fitting the material response using Fourier theory can only be accepted as approximate. As shown in section \ref{discussion},  the requirement of long times is particularly difficult to satisfy even if lengthscale corrections are applied. In general, in situations where the requirements for Fourier behavior are not carefully met, the effective thermal conductivity obtained by such fits will depend on the fitting time window, the measurement location \cite{lingping2014}, and the experimental geometry, highlighting the fact that it is not a material property and casting doubt over the reliability of the inverse calculation.

To avoid these conceptual but also practical problems, in the present work, we propose a methodology for reconstructing the relaxation time distribution in the material from thermal spectroscopy experimental results using an approach that at no time makes an assumption of an underlying Fourier behavior. In our work, the reconstruction problem is formulated as a non-convex optimization problem whose goal is to minimize the difference between the experimental material response and the response as calculated by a BTE-based model of the experimental process. Here, we note the recent paper \cite{Minnich2015} in which reconstruction is based on a Fourier-space solution of a BTE model of the two-dimensional transient thermal grating experiment. Despite emphasizing the BTE more than in previous approaches, the work in \cite{Minnich2015}, ultimately, still reverts to a Fourier-based formulation by fitting effective thermal diffusivities and introducing related approximations (e.g. late times compared to the mean phonon relaxation time). 

A further distinguishing feature of our work is the use of deviational Monte Carlo (MC) simulation \cite{PRB2011, APL2012, ARHT, PRB2015} as a method to solve the BTE which expands the domain of applicability of the approach, since it does not rely on the simplicity of the experimental setup or the degree to which it is amenable to analytical treatment.

The remainder of the paper is organized as follows. In section \ref{formulation}, we formulate the reconstruction as an optimization problem requiring only solutions of the BTE and present the optimization framework used for the reconstruction. In section \ref{Application/validation}, we discuss the implementation of the proposed method in the context of an archetypal experimental setup, namely that of a transient thermal grating \cite{Minnich2015, Hua2014, Collins2013}. In the same section, we validate our proposed method using synthetic data generated through Monte Carlo (MC) simulation. In section \ref{discussion} we provide some comparison to effective thermal conductivity approaches and a discussion on the validity of the latter. In section \ref{conclusion} we provide a summary of our work and suggestions for future improvement.

\section{Formulation} 
\label{formulation}

\subsection{Governing equations}
\label{Governing_equ}
Given the small temperature differences usually associated with thermal spectroscopy experiments, here we consider the linearized BTE 
\begin{equation} \label{BTE}
\frac{ \partial e^d } { \partial t } + \textbf{v}_{\omega} \cdot \nabla_{\textbf{x}} e^d = -\frac{ e^d- (de^{eq}/dT)_{T_{eq}} \Delta \widetilde{T} } { \tau_{\omega} } ,
\end{equation}
where $e^d= e^d (t, \mathbf{x}, \omega, \mathbf{\Omega})= e-e^{eq}_{T_{eq}}= \hbar \omega (f- f^{eq}_{T_{eq}}) $ is the deviational energy distribution, $\omega$ is the phonon frequency, $\mathbf{\Omega}$ is the phonon traveling direction, $f= f (t, \mathbf{x}, \omega, \mathbf{\Omega})$ is the occupation number of phonon modes, $\textbf{v}_{\omega}= \textbf{v} (\omega)$ is the phonon group velocity, $\tau_{\omega}= \tau (\omega)$ is the frequency-dependent relaxation time and $\hbar$ is the reduced Planck constant. Here and in what follows, we use $\omega$ to denote the dependence on both frequency and polarization.

The above equation is linearized about the equilibrium temperature $T_{eq}$, to be understood here as the experimental reference temperature. In general, $\tau_{\omega}= \tau (\omega, T)$; however, as a result of the linearization, $\tau_{\omega}= \tau (\omega, T_{eq}) \equiv \tau (\omega)$; in other words, the solutions (and associated reconstruction) are valid for the experiment baseline temperature $T_{eq}$. Also, $(de^{eq}/dT)_{T_{eq}}= \hbar \omega ( df^{eq}_T/dT ) \vert_{T_{eq}}$ and $f^{eq}_{T}$ is the Bose-Einstein distribution with temperature parameter $T$, given by
\begin{equation} \label{BE}
f^{eq}_{T} (\omega) = \frac{1} {\exp (\hbar \omega/ k_B T)- 1} ,
\end{equation} 
where $k_B$ is Boltzmann's constant. Finally, $\Delta \widetilde{T}= \Delta \widetilde{T} (t, \mathbf{x})= \widetilde{T}- T_{eq}$ is referred to as the deviational pseudo-temperature ($\widetilde{T} (t, \mathbf{x} )$ is the pseudo-temperature). Note that the deviational pseudo-temperature, which is different from the deviational temperature defined below, is defined by the energy conservation statement \cite{ARHT}
\begin{equation}
\int_{\mathbf{\Omega}} \int_{\omega} \left[ \frac{C_\omega} {\tau_{\omega}} \Delta \widetilde{T} - \frac{e^d} {\tau_{\omega}} D_{\omega} \right] d \omega d \mathbf{\Omega}= 0 , \label{pseudotemperature}
\end{equation} 
in which $D_{\omega}= D(\omega)$ is the density of states, $C_{\omega}= C (\omega; T_{eq})= D_{\omega} (de^{eq}/dT)_{T_{eq}}$ is the frequency-dependent volumetric heat capacity, and $d \mathbf{\Omega}= \sin (\theta) d \theta d \phi$ represents the differential solid angle element such that $\theta$ and $\phi$ refer to the polar and azimuthal angles in the spherical coordinate system, respectively. The temperature $T (t, \mathbf{x})$ is computed from 
\begin{equation}
\int_{\mathbf{\Omega}} \int_{\omega} \left[ C_{\omega} \Delta T- e^d D_{\omega} \right] d \omega d \mathbf{\Omega}= 0 , \label{temperature}
\end{equation}
where $\Delta T (t, \mathbf{x})= T- T_{eq}$ is the deviational temperature. The frequency-dependent free path is given by 
\begin{equation} 
\Lambda_{\omega}= v_{\omega} \tau_{\omega} ,
\label{lambda}
\end{equation} 
where $v_{\omega}= || \mathbf{v}_{\omega} ||$ is the group velocity magnitude.

\subsection{Inverse problem formulation} 
\label{InverseProblem}
Our goal is to obtain an accurate and reliable approximation to the function $\tau_{\omega}$ from the experimental measurements, with the latter typically in the form of a temperature (relaxation) profile. Reconstruction of the free path distribution follows from relationship \eqref{lambda}; in other words, the group velocities $v_{\omega}$ are assumed known, since they can be reliably calculated either experimentally \cite{GrapheneRaman, SiCRaman, ZincXRay, CarbonNanotubeDispersion} by means of Raman spectroscopy, x-ray scattering, etc, or theoretically \cite{RhodiumDispersion, GaAsDispersion, GraphiteDispersion, AlDispersion} using methods such as density functional theory (DFT).

We propose the use of an optimization framework in which the experimental measurements provide targets that need to be reproduced by the BTE solution; in other words, $\tau_{\omega}$ is determined as the function that optimizes (minimizes) the discrepancy between the experimental result and the BTE prediction for the same quantity. 
One important consideration is use of a suitable discrete representation for $\tau_{\omega}$. In our formulation any number of longitudinal/transverse acoustic/optical branches $\tau_{\omega}^S$ may exist and are solved for explicitly, where the superscript $S$ denotes the branch; the total scattering rate $\tau_{\omega}^{-1}$ is obtained using the Matthiessen's rule. In the examples following below we have taken
\begin{equation}
\tau^{-1}_{\omega} = \left( \tau^{LA}_{\omega} \right)^{-1}+ \left( \tau^{TA_1}_{\omega} \right)^{-1}+ \left( \tau^{TA_2}_{\omega} \right)^{-1} ,
\label{mixingrule}
\end{equation}
where $LA$ denotes the Longitudinal Acoustic branch, while $TA_1$ and $TA_2$ represent the two Transverse Acoustic branches. We have not considered optical phonons in this work because acoustic modes account for most of the heat conduction \cite{LatticeDynamics2011}, but also because the proposed methodology can be straightforwardly extended to include them.

In order to account for the most general situation where no a-priori information is available on the functional form of the $\tau_{\omega}= \tau(\omega)$ relation, we have used a piecewise linear relation between $\log (\tau_{\omega}^S)$ and $\log (\omega)$. The intersections between the piecewise linear segments are smoothed via a third order polynomial function (of $\log (\omega)$) that guarantees continuity of the relaxation time function and its first derivative \cite{smoothing} (the continuity requirement may be easily removed, allowing jumps between different segments). Note that our parametrization is equivalent to a (piecewise) relationship of the form of $\tau_{\omega}= c_1 \omega^{c_2}$ for some real numbers $c_1$ and $c_2$, which is related to relations commonly used in the literature \cite{LatticeDynamics2011, Holland}. Our approach, however, does not constrain the value of $c_2$ to any particular value or even an integer value. The actual functional form is given by 
\begin{multline}
\log \left( \tau^{S}_{\omega} \right) =\\ \sum_{j=0}^{M-1} \left[ \frac{ \log \left( \tau^{S}_{\omega_{j+1}} \right) - \log \left( \tau^{S}_{\omega_j} \right) } {\log \left( \omega^{S}_{j+1} \right)- \log \left( \omega^{S}_{j} \right) } \left( \log(\omega)- \log \left( \omega^{S}_{j} \right) \right)+ \log \left( \tau^{S}_{\omega_j} \right) \right] \mathds{1}_{ \omega \in [X^{S}_{2j}, X^{S}_{2j+1}]} +\\ \sum_{j=1}^{M-1} \left[ a^{S}_{j} \left[ \log(\omega) \right]^3+ b^{S}_{j} \left[ \log(\omega) \right]^2+ c^{S}_{j} \log(\omega)+ d^{S}_{j} \right] \mathds{1}_{\omega \in [X^{S}_{2j-1}, X^{S}_{2j}]} , \label{parametrization_2TA} 
\end{multline}
where $M$ determines the number of segments. We note that since the minimum and maximum frequencies for each branch ($\omega^{S}_{0}$ and $\omega^{S}_{M}$) are known (input), there are $2M$ unknowns in the model for each branch $S \in \{LA, TA_1, TA_2 \}$, consisting of $\omega^{S}_{1}$,..., $\omega^{S}_{M-1}$, and $\log \left( \tau^{S}_{\omega_0} \right)$,..., $\log \left( \tau^{S}_{\omega_M} \right)$, which are the inputs to the optimization algorithm, leading to a total of $6M$ unknowns (see section \ref{OptimAlgorithm} for implementation details; also note the input unknowns are denoted ${\bf p}^k$ in that section). Also, $\mathds{1}_{\omega \in \mathcal{S}}$ denotes the indicator function whose value is $1$ if $\omega \in \mathcal{S}$, and $0$ if $\omega \not\in \mathcal{S}$. 

Once $\omega^{S}_{j}$s and $\log \left( \tau^{S}_{\omega_j} \right)$s are known, the coefficients $a^{S}_{j}$, $b^{S}_{j}$, $c^{S}_{j}$, $d^{S}_{j}$, and the intervals in which each line or polynomial function is active inside, $X^{S}_{j}$s, can be computed. The relationships for calculating these parameters are provided in Appendix A. 

In the interest of compactness, we will use the vectorial notations $\pmb{\tau}^{S}= \left( \tau^{S}_{\omega_0},..., \tau^{S}_{\omega_M} \right)$ and $\pmb{\omega}^{S}= \left( \omega^{S}_{1},..., \omega^{S}_{M-1} \right)$ to represent the unknown parameters. Moreover, we will use the symbol ${\bf U}$ to denote the set of all unknown vectors.

The reconstruction proceeds by minimizing the objective function
\begin{equation}
\mathcal{L}= \min_{{\bf U}} \left[ \frac{ \sum_{t, \textbf{x}, L} | T_{m} (t, \mathbf{x}; L)- T_{BTE} (t, \mathbf{x}; L, {\bf U}) | } {N}+ \alpha \Bigg| 1- \frac{1} {3\kappa} \int_{\omega} C_{\omega} \tau_{\omega} ({\bf U}) v^2_{\omega} d \omega \Bigg| \right] , \label{objectivefunction_2TA} 
\end{equation}
where $T_m (t, \mathbf{x}; L)$ denotes the experimentally measured temperature, $T_{BTE}$ is the temperature computed from solution of the BTE, and $N$ is the total number of (independent) measurements available. As indicated above, $T_m$ is in general a function of space, time, but also the characteristic lengthscale of the thermal relaxation problem, $L$. As a result, the optimization process can be based on data for $T_{BTE}$ and $T_m$ at various time instances, spatial locations, and for different characteristic system parameters, with $\sum_{t, \textbf{x}, L} 1= N$. We also point out that $T_{BTE}$ may be obtained by any method that can provide accurate solutions of \eqref{BTE} as applied to the experimental setup; here, we consider both semi-analytical solution (using Fourier transform techniques) and MC simulation.

The second term in \eqref{objectivefunction_2TA} exploits the fact that the bulk value of the thermal conductivity, $\kappa$, is usually known, to enhance the importance of the low frequency modes in the optimization. Although this term is ``optional", we have found that including this term with a weight of $0.01< \alpha< 1$ improves the reconstruction quality considerably in the low frequency regime, because the low density of states associated with those frequencies prevents them from influencing the optimization process if the objective function only includes a comparison between $T_{BTE}$ and $T_m$ (the first term in \eqref{objectivefunction_2TA}).

\subsection{Optimization algorithm} 
\label{OptimAlgorithm}
Determination of ${\bf U}$ (the vectors $\pmb{\tau}^{LA}$, $\pmb{\tau}^{TA_1}$, $\pmb{\tau}^{TA_2}$, $\pmb{\omega}^{LA}$, $\pmb{\omega}^{TA_1}$, and $\pmb{\omega}^{TA_2}$) which minimize the objective function \eqref{objectivefunction_2TA} is achieved using the Nelder-Mead (NM) algorithm \cite{NelderMead}, a simplex-based search method that is free from gradient computation. Although NM can neither be categorized as a local optimizer or a global optimizer, it performs significantly better than common local minimizers (which fail to converge to the correct solution in the presence of highly non-convex objective functions such as \eqref{objectivefunction_2TA}) \cite{NM2014}, while it is significantly less costly than common global optimizers like genetic algorithms \cite{GeneticAlgorithm} or simulated annealing algorithms \cite{SAAlgorithm}. 

\begin{figure} [H]
\begin{subfigure}{.5\textwidth}
\centering
\includegraphics[width=0.7\linewidth]{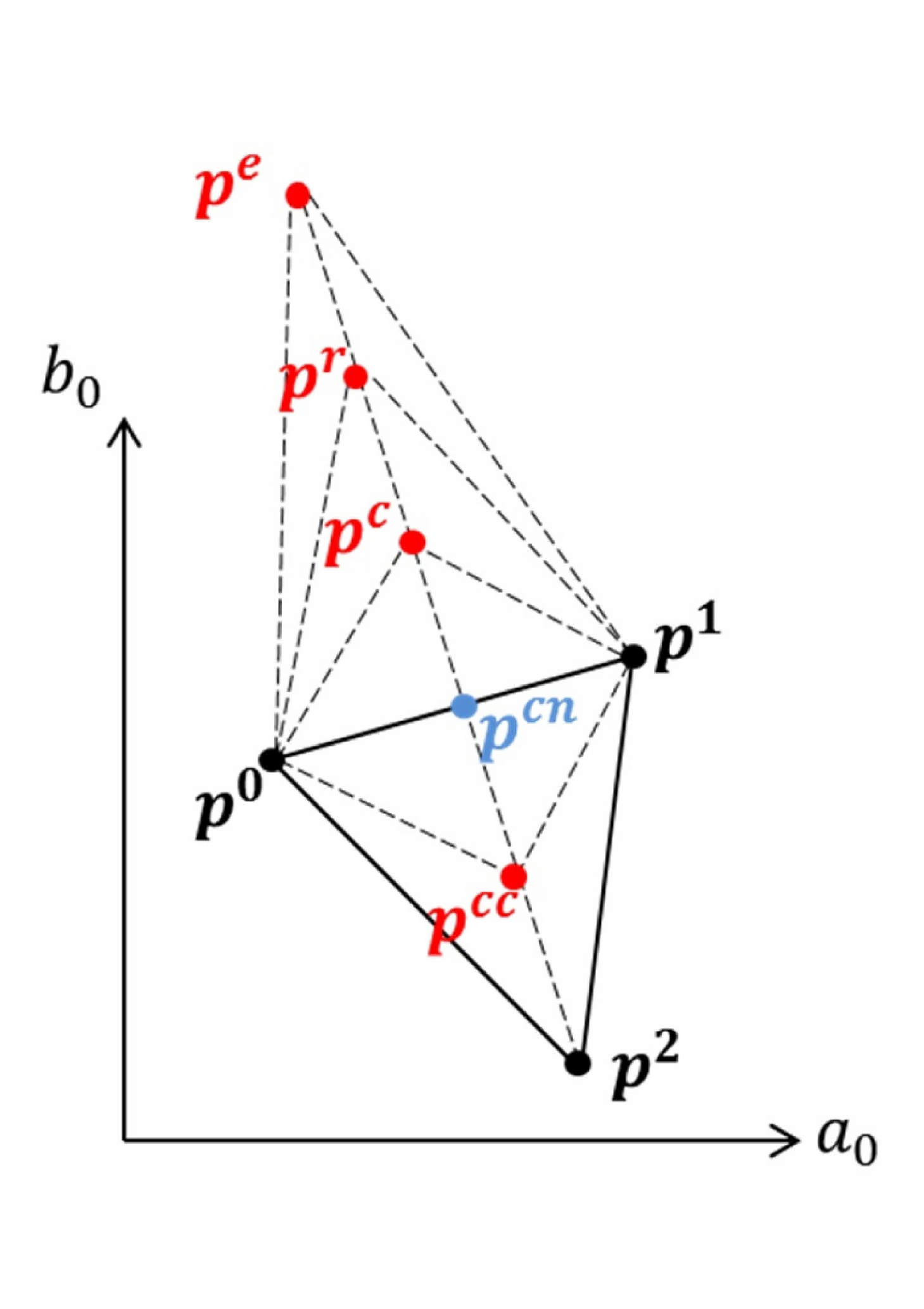}
\caption{}
\label{NM1}
\end{subfigure}
\begin{subfigure}{.5\textwidth}
\centering
\includegraphics[width=0.7\linewidth]{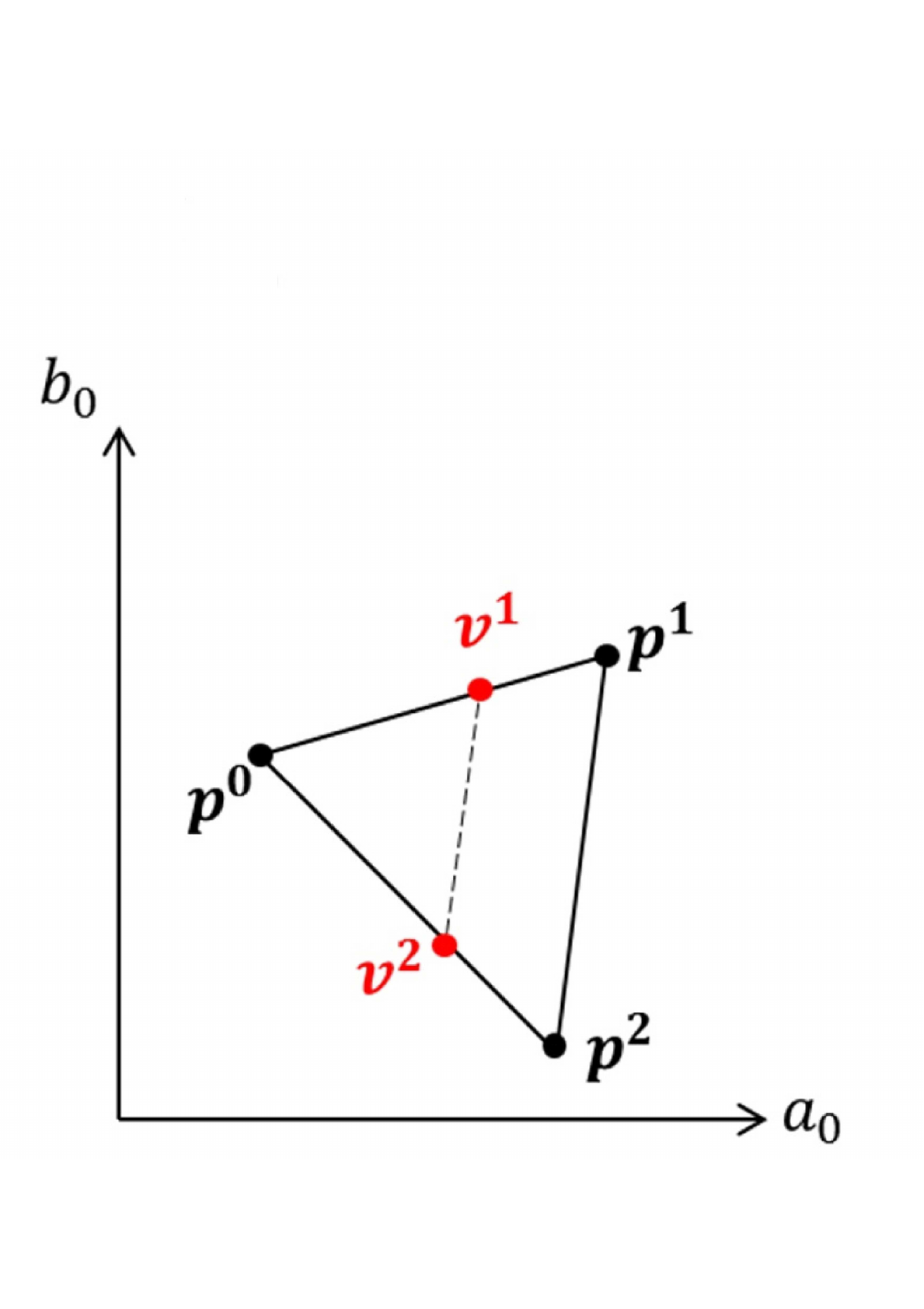}
\caption{}
\label{NM2}
\end{subfigure}
\caption{An example of simplex update in the NM algorithm; the case of two unknowns (corresponding to $n=2$) is shown. (a)- Point $\mathbf{p}^2$ of the previous simplex (which consists of points $\mathbf{p}^0$, $\mathbf{p}^1$, and $\mathbf{p}^2$) is replaced with a new point from the set $\{ \mathbf{p}^c, \mathbf{p}^{cc}, \mathbf{p}^r, \mathbf{p}^e \}$. The new simplex consists of $\mathbf{p}^0$, $\mathbf{p}^1$, and the replacement of $\mathbf{p}^2$. (b)- If none of the candidate points in figure \ref{NM1} have an objective function smaller than $\mathcal{L} (\mathbf{p}^2)$, then the simplex shrinks. The new simplex is $\mathbf{p}^0$, $\mathbf{v}^1$, and $\mathbf{v}^2$. }
\label{NM}
\end{figure}

In the NM algorithm, the optimization process proceeds as follows (adopted from \cite{NMLagarias}): 
\begin{enumerate}
\item Start with $n+1$ initial points $\mathbf{p}^k$, $k= 0,..., n$, in the $n$-dimensional space defined by the $n$ unknown parameters (here $n=6M$). A possible choice of initial simplex (the set of $n+1$ initial points) is $\mathbf{p}^0$ and $\mathbf{p}^k= \mathbf{p}^0+ \delta_k \mathbf{e}_k$ for scalars $\delta_k$, $k= 1,..., n$; here $\mathbf{p}^0$ is an arbitrary point in the $n$-dimensional space of unknown parameters and $\mathbf{e}_k$ is the unit vector in the $k$-th direction \cite{NMPress}. 
\item Order the points $\mathbf{p}^k$, $k= 0,..., n$, in an ascending order based on their objective function values. Assign new superscripts such that $\mathcal{L} (\mathbf{p}^0) \leq ... \leq \mathcal{L}(\mathbf{p}^{n-1}) \leq \mathcal{L} (\mathbf{p}^n)$.
\item Calculate the centroid of the $n$ points with lowest objective function values, $\mathbf{p}^{cn}= \frac{1} {n} \sum_{k= 0}^{n-1} \mathbf{p}^k$.
\item Calculate the reflected point, $\mathbf{p}^r= \mathbf{p}^{cn}+ \mu \left( \mathbf{p}^{cn}- \mathbf{p}^n \right)$, where $\mu > 0$. 
\item If $\mathcal{L} (\mathbf{p}^0) \leq \mathcal{L} (\mathbf{p}^r) < \mathcal{L} (\mathbf{p}^{n-1})$, replace $\mathbf{p}^n$ with $\mathbf{p}^r$ and start the next iteration; go to step 2. 
\item If $\mathcal{L} (\mathbf{p}^r) < \mathcal{L} (\mathbf{p}^0)$, compute $\mathbf{p}^e= \mathbf{p}^r+ \gamma \left(\mathbf{p}^r- \mathbf{p}^{cn} \right)$, the expanded point, where $\gamma > 0$.
\begin{enumerate}
\item If $\mathcal{L} (\mathbf{p}^e) < \mathcal{L} (\mathbf{p}^r)$, replace $\mathbf{p}^n$ with $\mathbf{p}^e$ and start the next iteration; go to step 2.
\item If $\mathcal{L} (\mathbf{p}^e) \geq \mathcal{L} (\mathbf{p}^r)$, replace $\mathbf{p}^n$ with $\mathbf{p}^r$ and start the next iteration; go to step 2.
\end{enumerate} 
\item If $\mathcal{L} (\mathbf{p}^r) \geq \mathcal{L} (\mathbf{p}^{n-1})$
\begin{enumerate}
\item If $\mathcal{L} (\mathbf{p}^r) < \mathcal{L} (\mathbf{p}^n)$, compute $\mathbf{p}^c= \mathbf{p}^{cn}+ \rho \left( \mathbf{p}^r- \mathbf{p}^{cn} \right)$, the contracted point, where $0< \rho< 1$. If $\mathcal{L} (\mathbf{p}^c) \leq \mathcal{L} (\mathbf{p}^r)$, replace $\mathbf{p}^n$ with $\mathbf{p}^c$ and start the next iteration; go to step 2. Otherwise, go to step 8.
\item If $\mathcal{L} (\mathbf{p}^r) \geq \mathcal{L} (\mathbf{p}^n)$, compute $\mathbf{p}^{cc}= \mathbf{p}^{cn}+ \rho \left( \mathbf{p}^n- \mathbf{p}^{cn} \right)$, the new contracted point. If $\mathcal{L} (\mathbf{p}^{cc}) \leq \mathcal{L} (\mathbf{p}^n)$, replace $\mathbf{p}^n$ with $\mathbf{p}^{cc}$ and start the next iteration; go to step 2. Otherwise, go to step 8.
\end{enumerate}
\item For $k= 1,..., n$, compute $\mathbf{v}^k= \mathbf{p}^0+ \sigma \left( \mathbf{p}^k- \mathbf{p}^0 \right)$ where $0< \sigma< 1$, replace $\mathbf{p}^k$ with $\mathbf{v}^k$, and start the next iteration (go to step 2). 
\end{enumerate}
In words, in each iteration, the objective is to eliminate the worst point (with the largest value of objective function) among the current $n+1$ points of the simplex via reflection (step 5 or 6-b), expansion (step 6-a), contraction (step 7) or shrinkage (step 8). Figure \ref{NM} illustrates this algorithm for the case $n=2$. In the present work, the parameters $\mu, \gamma$, and $\rho$ are taken to be 1, 1, and 0.5, respectively, as proposed in \cite{NelderMead, NMLagarias, NMimprovement}. We have also chosen the shrinkage coefficient to be $\sigma= 0.9$ as recommended in \cite{NMimprovement} for robustness in the presence of noise.

The most expensive step in this process is the calculation of the objective function \eqref{objectivefunction_2TA} which requires evaluation of $T_{BTE}$. The NM algorithm typically requires one or two evaluations of \eqref{objectivefunction_2TA} at each iteration, except when a shrinkage step takes place where $6M+2$ evaluations of \eqref{objectivefunction_2TA} are required. Single function evaluation occurs as a result of successful initial reflection corresponding to step 5 of the algorithm. Two evaluations of the objective function result from either expansion at step 6-a, reflection at step 6-b, outside contraction at step 7-a or inside contraction at step 7-b. The simplex typically starts to shrink near the convergence point where it cannot find any new direction that yields lower values of the objective function and consequently instead searches for such point inside the previous simplex by shrinking it toward its best vertex \cite{NMimprovement}.

\section{Application example and validation} 
\label{Application/validation}
In this section, we consider the one-dimensional transient thermal grating (TTG) experiment as an example for illustrating and validating the methodology presented in section \ref{formulation}. The mathematical model of the TTG experiment will be briefly discussed in section \ref{TTG}. Using this mathematical formulation, we generate synthetic experimental data; this process is described in section \ref{synthetic}. We then use these synthetic data to reconstruct $\tau_{\omega}$ for the material model used in the MC simulations; the details of the reconstruction process are discussed in section \ref{example:reconstruction}. We consider reconstruction both via semi-analytical solution of the BTE and MC simulation; the advantages of each approach are discussed and contrasted. In section \ref{validation}, we compare the reconstructed relaxation times and free path distribution with those of the original data used in the MC simulations for generating the synthetic data.

\subsection{The TTG experiment} 
\label{TTG}
TTG experiment has been widely used to study transient thermal transport. In this experiment, a sinusoidal profile with unit amplitude is generated by crossing two short pump pulses \cite{Rogers1994}. By assuming a one-dimensional heat transport along the $x$ direction, the BTE from equation \eqref{BTE} can be written in the form 
\begin{equation} \label{BTE_TTG}
\frac{\partial e^d} {\partial t} + v_{\omega} \cos(\theta) \frac{\partial e^d} {\partial x}= -\frac{e^d- (de^{eq}/dT)_{T_{eq}} \Delta \widetilde{T}} {\tau_{\omega}}+ (de^{eq}/dT)_{T_{eq}} \delta (t) \mathrm{e}^{ 2 \pi i x/L} ,
\end{equation}
in which the last term is due to the sinusoidal initial temperature profile. Here, $L$ is the grating wavelength (see discussion in section \ref{InverseProblem}), and $i$ is the imaginary unit. This equation is sufficiently tractable to allow analytical solutions in Fourier/reciprocal space \cite{Hua2014, Collins2013}. We follow a similar procedure (see Appendix B) to obtain a Fourier space solution for $T_{BTE}$; the final result is 
\begin{equation}
\Delta {T}_\zeta= \frac{ \mathrm{e}^{2 \pi i x/L}} {C} \left[ \int_{\omega} S_{\omega} \tau_{\omega}^2 d \omega+ \frac{ \left( \int_{\omega} S_{\omega} \tau_{\omega} d \omega \right)^2 } {\int_{\omega} \left[ C_{\omega} \tau_{\omega}^{-1}- S_{\omega} \right] d \omega} \right] , \label{Analytical}
\end{equation}
where $C= \int_{\omega} C_{\omega} d \omega$ is the volumetric heat capacity and $\Delta T_{\zeta}= \Delta T (\zeta, x)$ denotes the Fourier transform of $\Delta T (t, x)$ with respect to the time variable. Also, $S_{\omega}= S (\omega, \zeta)$ is given by
\begin{equation}
S_{\omega}= \frac{ i C_{\omega} L} { 4 \pi v_{\omega} \tau_{\omega}^2 } \ln \left( \frac{\tau_{\omega} \zeta- v_{\omega} \tau_{\omega} 2 \pi L^{-1}- i} {\tau_{\omega} \zeta+ v_{\omega} \tau_{\omega} 2 \pi L^{-1}- i} \right) . \label{S_omega}
\end{equation}
We note here that the temperature profile required for evaluating \eqref{objectivefunction_2TA} is given by $T_{BTE} (t, x)=  \Delta {T}(t, x)+ T_{eq}$. 

\subsection{Generation of synthetic experimental data}
\label{synthetic}
Instead of using experimental data (for $T_m$) to validate the proposed methodology, we use synthetic data generated from MC simulation of \eqref{BTE_TTG}. This approach enables considerably more precise validation because it sidesteps issues of experimental error, which is hard to quantify, but also modeling error (how accurately does \eqref{BTE_TTG} model the TTG experiment?), which is even harder to quantify. In other words, reconstruction from $T_m$ ``measurements" obtained from MC simulations should be able to reproduce the function $\tau_{\omega}$ used to generate the synthetic data exactly, making any discrepancies directly attributable to numerical/methodological error. 

In the interest of simplicity, the material was assumed to be silicon. However, in order to fully explore the ability of the reconstruction process to capture arbitrary functional relations $\tau_{\omega}= \tau (\omega)$, we consider both a simple analytical model and an ab initio model of this material. The simple model considered here is described in \cite{PRB2011, MinnichThesis} (thermal conductivity $\kappa= 143.8 \ Wm^{-1}K^{-1}$) and will be referred to as the \textit{Holland} model. The second model considered (thermal conductivity $\kappa= 139.7 \ Wm^{-1}K^{-1}$)  is described in \cite{Collins2013} and will be referred to as the \textit{ab initio} model throughout this paper.

Mirroring the experimental procedure as closely as possible, we obtain transient temperature relaxation data (solutions of \eqref{BTE_TTG}) at one spatial location for a number of characteristic (grating) lengths, namely $T_m (0 \leq t \leq t_L; L)$ ($= T_m (0 \leq t \leq t_L, x= 0; L)$ for example). These solutions were obtained using the adjoint Monte Carlo method recently proposed \cite{PRB2015} that is particularly efficient if solutions at particular spatial locations are of interest. The MC simulations used a sufficiently high number of particles ($\mathcal{N}_m= 10^9$) to ensure that the synthetic data was essentially noise free. As discussed in section \ref{validation}, we have also produced noisy synthetic data using significantly fewer particles in order to investigate the robustness of the optimization method to noise. Relaxation data was ``recorded" in the time period $0 \leq t \leq t_L$, where $t_L= \min (t_{L, 1\%}, 5\ ns)$; here, $t_{L, 1\%}$ denotes the time taken for the response to decay to 1\% of its original amplitude for each $L$. During this time period, 100 discrete $T_m$ measurements were obtained. Eight different wavelengths were simulated, namely, $L= 10 \ nm$, $50 \ nm$, $100 \ nm$, $500 \ nm$, $1 \ \mu m$, $5 \ \mu m$, $10 \ \mu m$, and $50 \ \mu m$. As a result, 800 total $T_m$ measurements were available for reconstruction ($N= 800$).

\subsection{Reconstruction}
\label{example:reconstruction}
Reconstruction proceeds by comparing solutions of 1-D TTG experiment, equation \eqref{BTE_TTG}, in the form of $T_{BTE} ( 0 \leq t \leq t_L; L, {\bf U}) $ to the counterpart $N$ measurements of $T_m$ as a means of finding the optimum unknown vectors. Here, solutions of \eqref{BTE_TTG} were obtained using inverse fast Fourier transform (IFFT) of \eqref{Analytical} as explained in \cite{FFT}, or by using adjoint MC simulations \cite{PRB2015}. Although MC simulations are more expensive, they are investigated here because they make the proposed method significantly more general, since they are not limited to problems which lend themselves to analytical or semi-analytical solutions. 

As discussed before, phonon group velocities were assumed known, 
while $\tau_{\omega}$ was described by the model given in \eqref{parametrization_2TA} with $M= 3$ (a piecewise linear function with three segments). As shown in section \ref{validation}, this approximation level gives very good results. 

To generate the initial starting points ($6M+1$ vertices of the simplex), we have used $\delta_k= 0.1 p^0_k$ as recommended in \cite{NMPress}, where $p^0_k$ represents the $k$-th component of vector $\mathbf{p}^0$ as discussed in the first step of the NM algorithm in section \ref{OptimAlgorithm}. Larger values of $\delta_k$ (we have tested $\delta_k= 10 p^0_k$) may lead to large steps during the initial iterations of the optimization, which consequently can move the simplex toward a wrong local minimum; on the other hand, very small values of $\delta_k$ (we have tested $\delta_k= 0.001 p^0_k$) lead to smaller steps throughout the optimization process, which require a (significantly) larger number of iterations in order to guarantee that the parameter space is adequately sampled. We have obtained similar performances to $\delta_k= 0.1 p^0_k$ using $\delta_k= 0.05 p^0_k$ and $\delta_k= 0.025 p^0_k$.

We perform the optimization in four stages. In the first stage, we solve for one segment ($M=1$) which is the same for all branches, leading to only two unknown parameters. Due to the small number of unknowns, this step is very cheap but very valuable since it results in a significant reduction in the value of objective function in only 10-20 iterations. In the second stage, $LA$ and $TA$ modes are still assumed to be the same but the intended number of segments is used, increasing the number of unknowns to $2M$ (six in the present case). The initial condition for this stage is taken to be the same as the optimized value of the previous stage (or a slightly perturbed version). In the third stage, we repeat the optimization process, now for $4M$ unknowns (the two $TA$ branches are assumed to be the same, $\tau_{\omega}^{TA_1}= \tau_{\omega}^{TA_2}$), starting from the optimized parameters of the second stage. Finally, we perform the optimization for all $6M$ unknowns starting from the optimized parameters of the previous stage (repeated for each of the three branches independently). This procedure is more robust than direct optimization for $6M$ unknowns, which depending on the initial condition may be trapped in a local minimum of \eqref{objectivefunction_2TA} characterized by significantly different relaxation times for the $LA$ and $TA$ branches.

To reduce the probability of the reconstruction being trapped in a local minimum, in our approach, the first and second stages were repeated starting from 5 distinct initial conditions. The result at each stage with the lowest value of $\cal{L}$ was used as the initial condition for the next stage. Although this consideration may not be necessary, it is expected to improve the quality of the optimization process (by providing a smaller final value of $\mathcal{L}$) due to the non-convexity of the objective function. 

Although the number of iterations varies depending on the initial condition, we have observed that for the value of $M$ used in our work (corresponding to $6M=18$ unknowns), the third and fourth stages require on the order of 150-200 iterations for convergence (each). 
When using analytical or semi-analytical (IFFT) solutions for $T_{BTE}$, the computational cost associated with each evaluation is very small and thus cost is not a consideration. Using MC simulation to obtain solutions for $T_{BTE}$ is more costly and is discussed below.

\subsubsection{Reconstruction using MC simulations} 
\label{Reconstruction_MC}
To reduce cost, MC simulations during the first two stages of the optimization process used $\mathcal{N}_{BTE}= 10^4$ particles for calculation of $T_{BTE}$. During the third and fourth stages of the process, the number of particles is increased to $\mathcal{N}_{BTE}= 10^6$.  

\begin{figure} [H]
\centering
\includegraphics[width=0.6\linewidth]{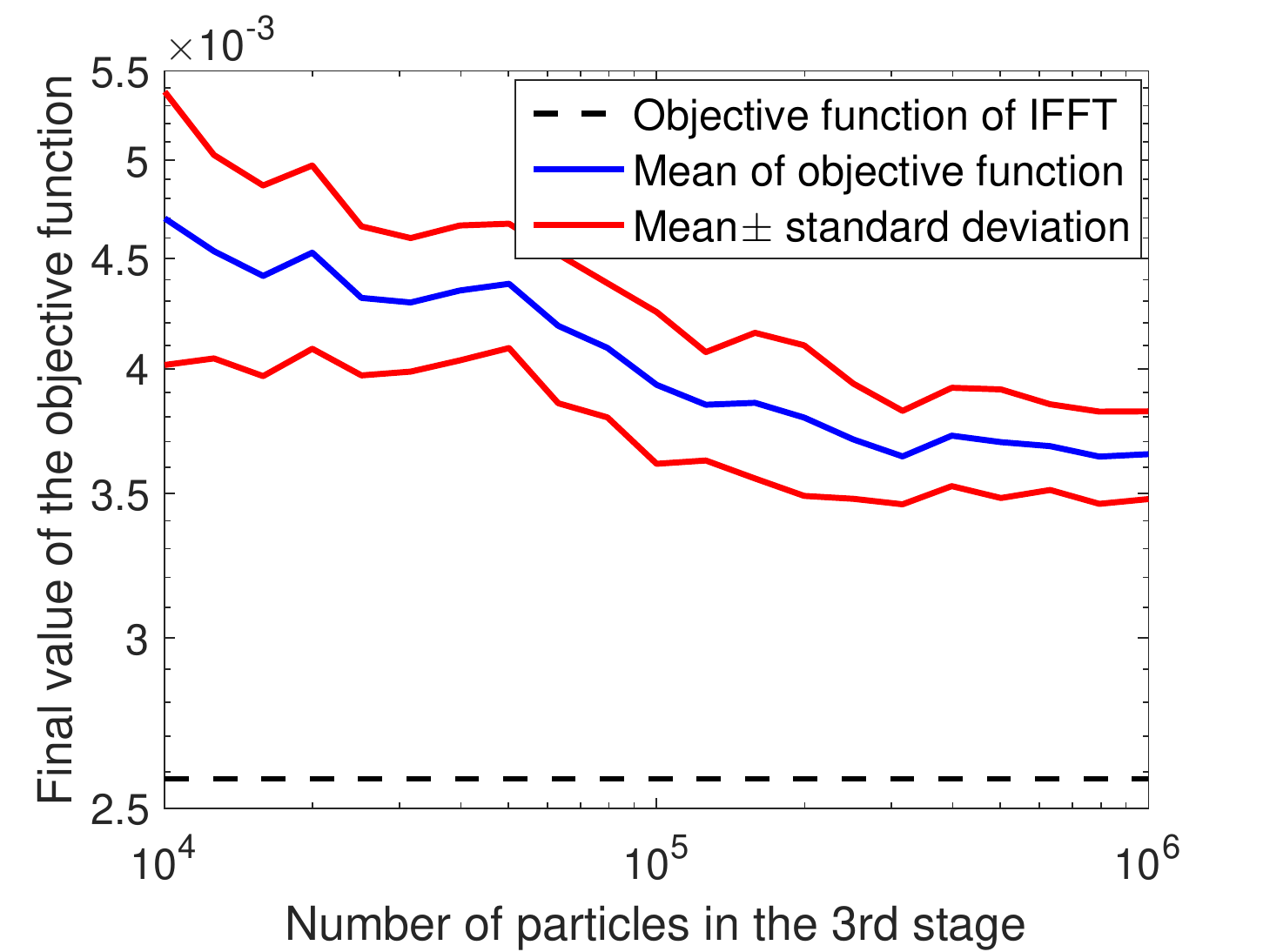}
\caption{Dependence of the final value of $\mathcal{L}$ at the third stage on $\mathcal{N}_{BTE}$. Increasing the number of particles decreases the mean and variance of $\mathcal{L}$.}
\label{obj_func_particle_num}
\end{figure}

Figure \ref{obj_func_particle_num} investigates the sensitivity of the final value of the objective function $\mathcal{L}$ in the third stage as a function of the number of particles used (in this stage of the optimization) for the ab initio material model. Specifically, the figure compares the mean and variance of the objective function compared to the third-stage final value obtained from IFFT of \eqref{Analytical} (which is free from noise and independent of number of particles). The MC result appears to converge (asymptotically) to the deterministic result with increasing number of computational particles, $\mathcal{N}_{BTE}$. Although this convergence appears to be slow, it is encouraging that the additional error (compared to deterministic reconstruction) is small and thus the shallow slope of the convergence implies that the reconstruction is quite robust to noise in the range $10^4 \leq \mathcal{N}_{BTE} \leq 10^6$.  

Since the computational time for each adjoint MC simulation is proportional to $\mathcal{N}_{BTE}$ \cite{PRB2015}, given our choices for $\mathcal{N}_{BTE}$ ($\mathcal{N}_{BTE}= 10^4$ for stages 1 and 2  and $\mathcal{N}_{BTE}= 10^6$ for stages 3 and 4), the cost of the first two stages of the optimization is negligible compared to the cost of the last two stages. Here we note that no attempt has been made to optimize this process; it is therefore possible that considerably more computational benefits may be possible if a more structured approach is used \cite{jcp99} that takes into account information such as that contained in figure \ref{obj_func_particle_num}. We also note that given the small cost of the first two optimization stages, repeating them a number of times starting from different initial conditions does not increase the cost of the reconstruction process significantly. 

For an average of 150-200 iterations, the cost of each of the third or fourth stages  corresponds to 300-600 MC simulations, considering that on average 2-3 evaluations of \eqref{objectivefunction_2TA} at each iteration are required. This number of iterations can be performed efficiently using the adjoint method proposed recently in \cite{PRB2015}. Note that a significant fraction of the computational effort is spent on iterations close to the final solution, which is primarily due to the high computational cost of the shrinkage step near convergence as was discussed previously in section \ref{OptimAlgorithm} (compare $6M+2= 20$ function evaluations associated with the shrinkage step, with one or two function evaluations associated with reflection, expansion or contraction). In contrast to the present case where validation required very small error tolerances, in actual practice, tolerances will be set by experimental considerations and are expected to be larger, reducing the number of expensive shrinkage processes and making the cost of reconstruction from experimental data smaller. 

\subsection{Validation}
\label{validation}
In this section, we discuss reconstruction results with particular emphasis on their ability to reproduce the original material properties ($\tau_{\omega}= \tau (\omega)$) used to generate the synthetic measurements $T_m$, referred to as ``true". Although our scheme returns $\tau_{\omega}$ as output, as is typical in the literature, we also provide comparison between the ``true" and ``reconstructed" cumulative distribution function (CDF) of free paths defined as $F (\Lambda)= \frac{1} {3 \kappa} \int_{\omega^*(\Lambda)} C_{\omega} v_{\omega}^2 \tau_{\omega} d \omega$, where $\omega^*(\Lambda)$ is the set of modes such that $\omega^*(\Lambda)= \{\omega | \Lambda_{\omega} \leq \Lambda\}$.

\subsubsection{Holland model}
\label{Holland}
In this section, we present results for the Holland material model. For this model, we assumed that the two $TA$ branches are the same and as a result, the optimization process consists of three stages with a total of $4M$ unknowns.

\begin{figure} [H]
\begin{subfigure}{.5\textwidth}
\centering
\includegraphics[width=0.98\linewidth]{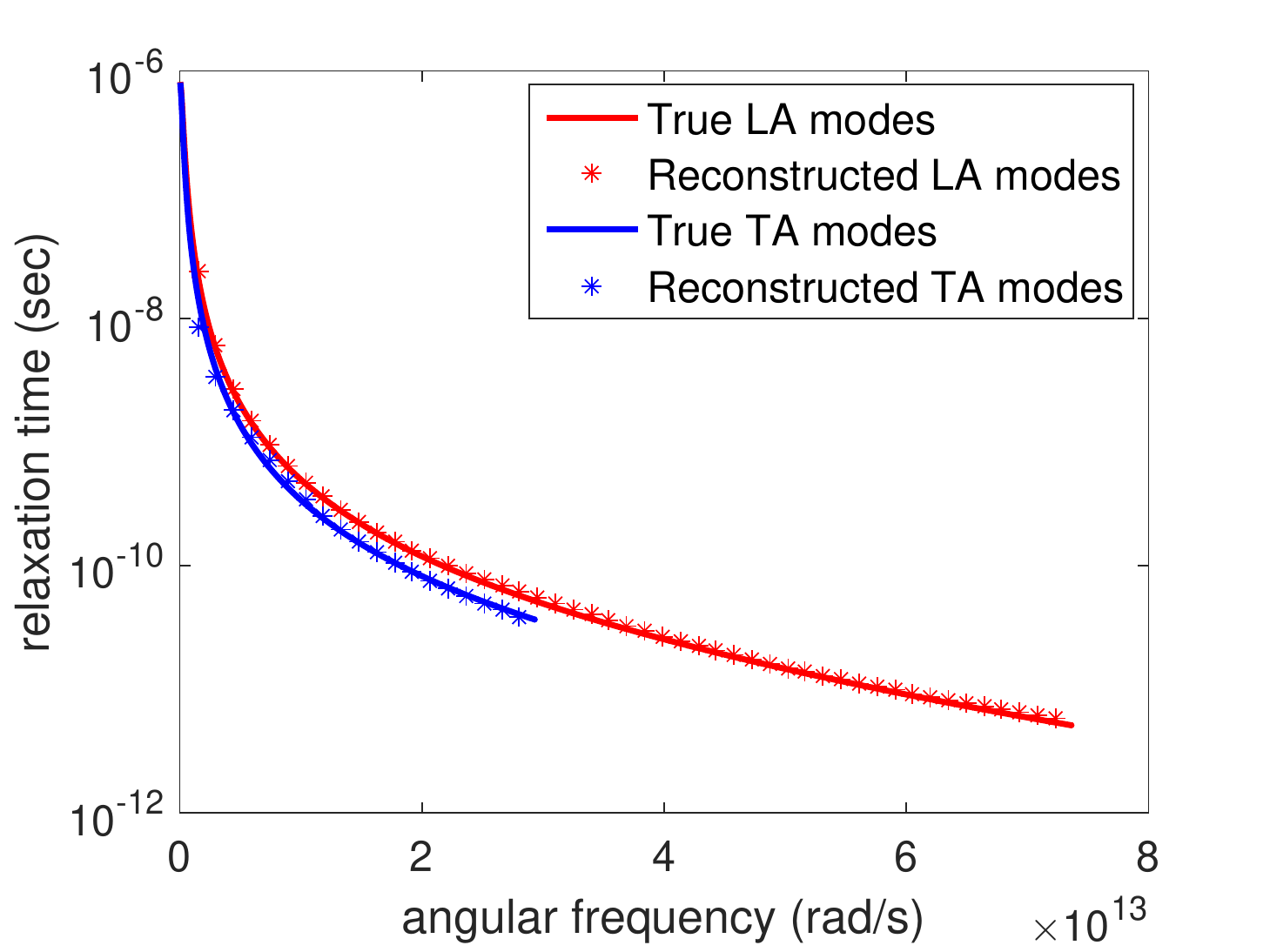}
\caption{relaxation times}
\label{}
\end{subfigure}
\begin{subfigure}{.5\textwidth}
\centering
\includegraphics[width=0.98\linewidth]{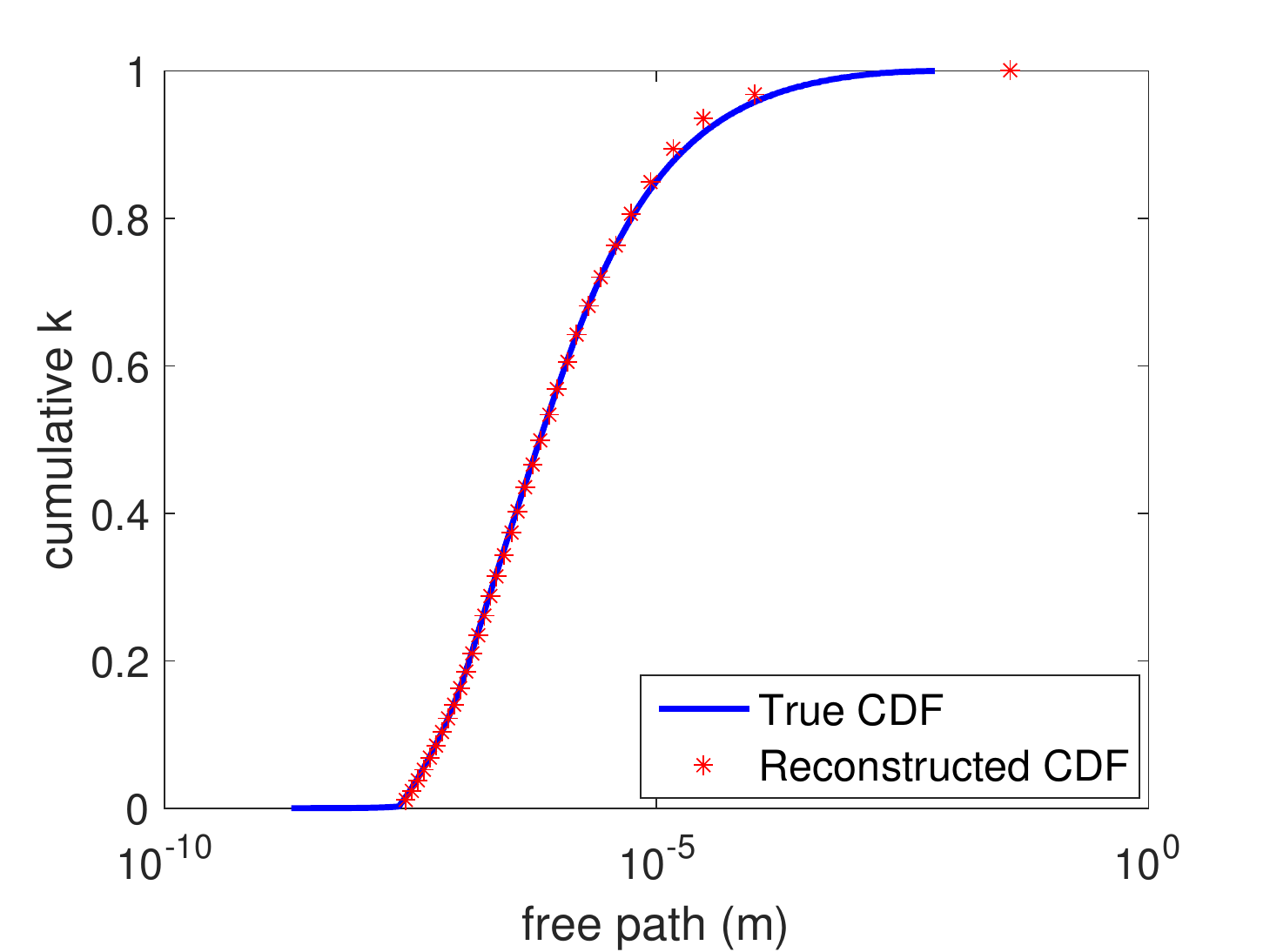}
\caption{free path distribution}
\label{}
\end{subfigure}
\caption{Comparison between true and reconstructed relaxation times and free path distribution using IFFT of the semi-analytical result \eqref{Analytical} for the Holland material model. }
\label{compwithanalytical}
\end{figure}

Figure \ref{compwithanalytical} shows a comparison between the true material parameters and the reconstructed ones using the semi-analytical solution \eqref{Analytical} for $T_{BTE}$. The agreement is excellent.

\begin{figure} [H]
\begin{subfigure}{.5\textwidth}
\centering
\includegraphics[width=1.0\linewidth]{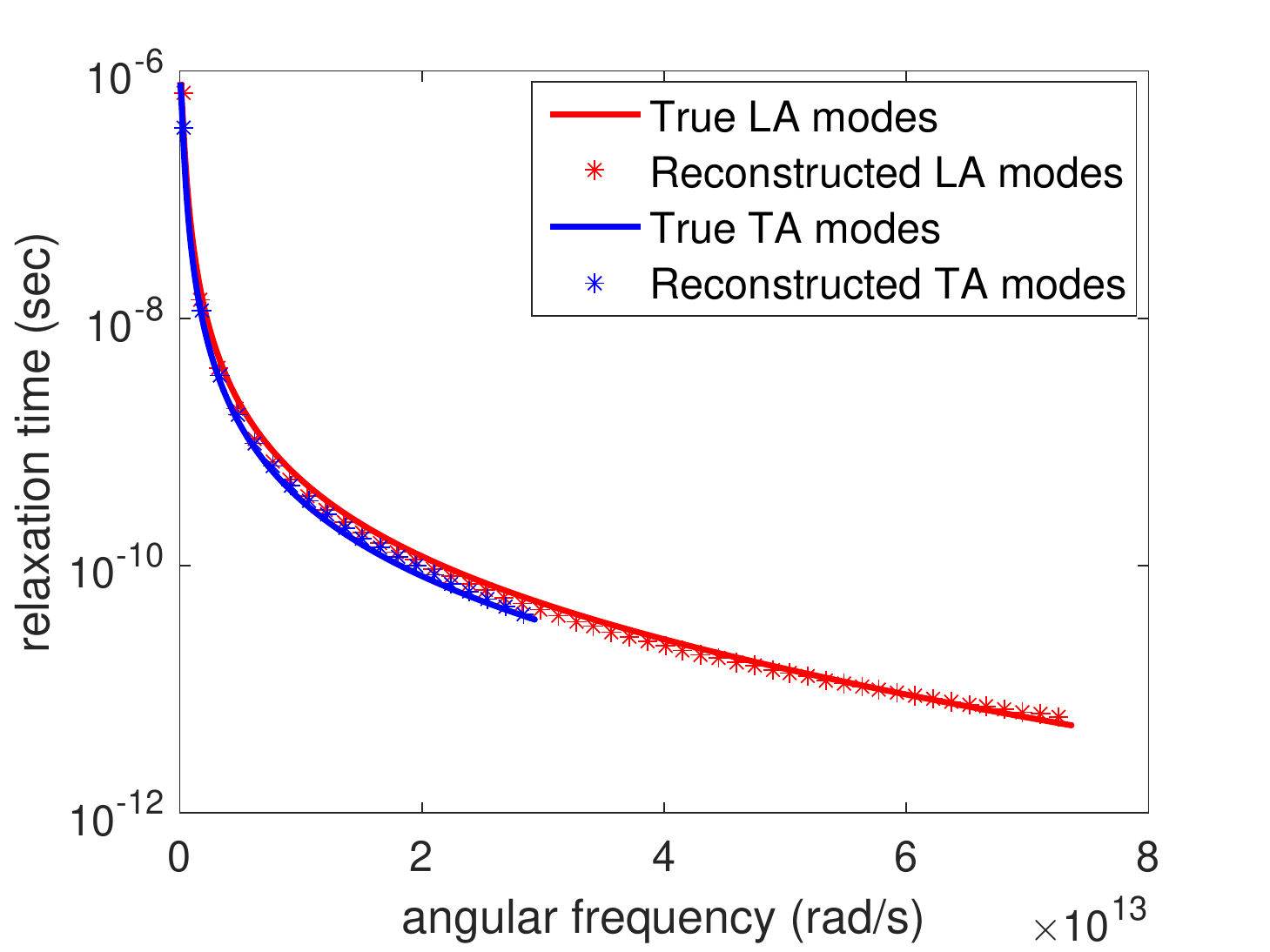}
\caption{relaxation times}
\label{}
\end{subfigure}
\begin{subfigure}{.5\textwidth}
\centering
\includegraphics[width=1.01\linewidth]{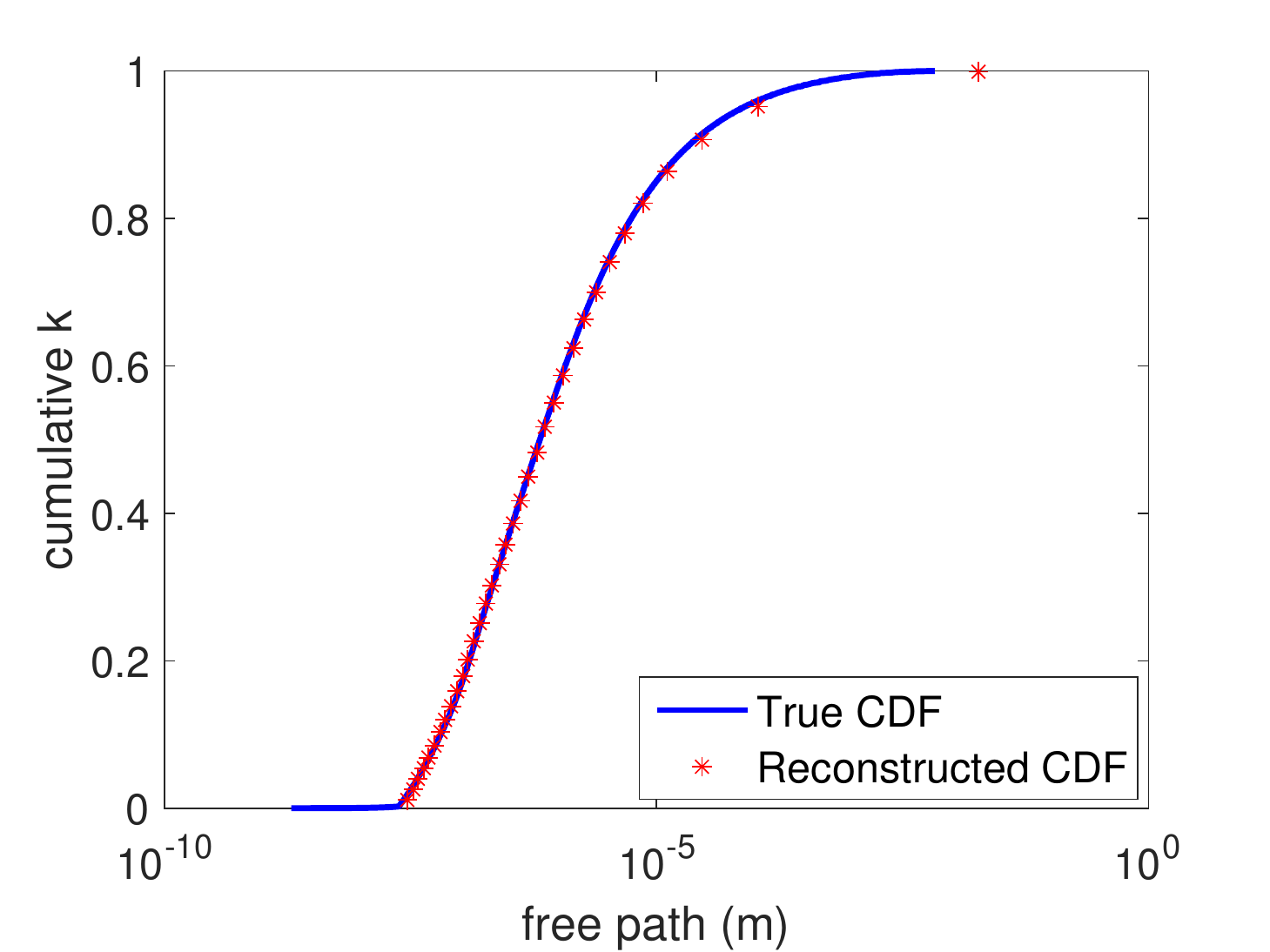}
\caption{free path distribution}
\label{}
\end{subfigure}
\caption{Comparison between true and reconstructed relaxation times and free path distribution using $T_{BTE}$ obtained from adjoint MC simulation using the Holland material model.}
\label{compwithMC}
\end{figure}

Figure \ref{compwithMC} shows a similar comparison for the case where  MC simulations were used for obtaining $T_{BTE}$. We observe that the error in reconstructed relaxation times is more significant when using MC simulations. We attribute this primarily to the noise associated with MC simulation and its interaction with the NM algorithm.  

\begin{figure} [H]
\begin{subfigure}{.5\textwidth}
\centering
\includegraphics[width=1.0\linewidth]{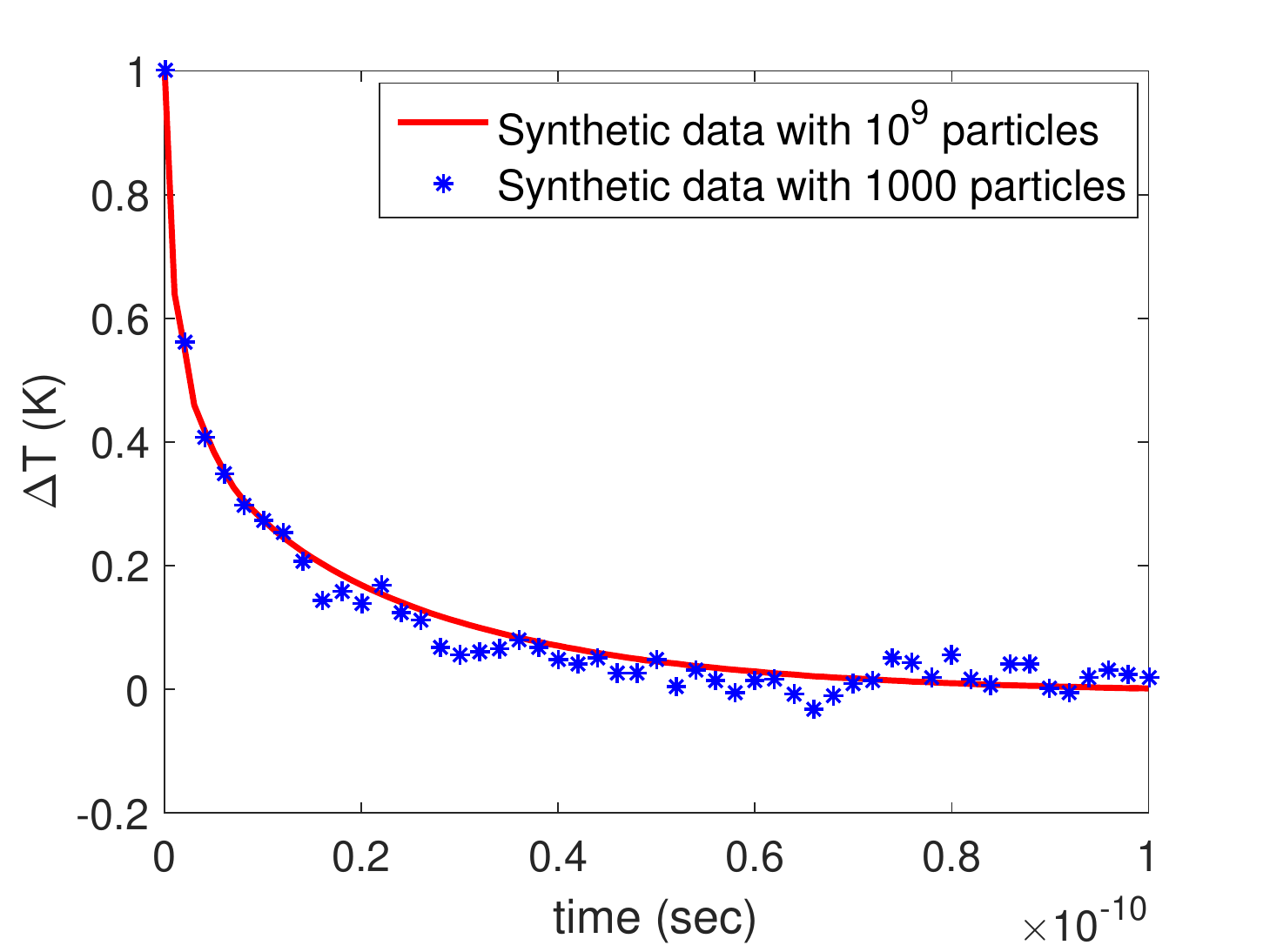}
\caption{temperature profile for $L= 10\ nm$}
\label{}
\end{subfigure}
\begin{subfigure}{.5\textwidth}
\centering
\includegraphics[width=1.0\linewidth]{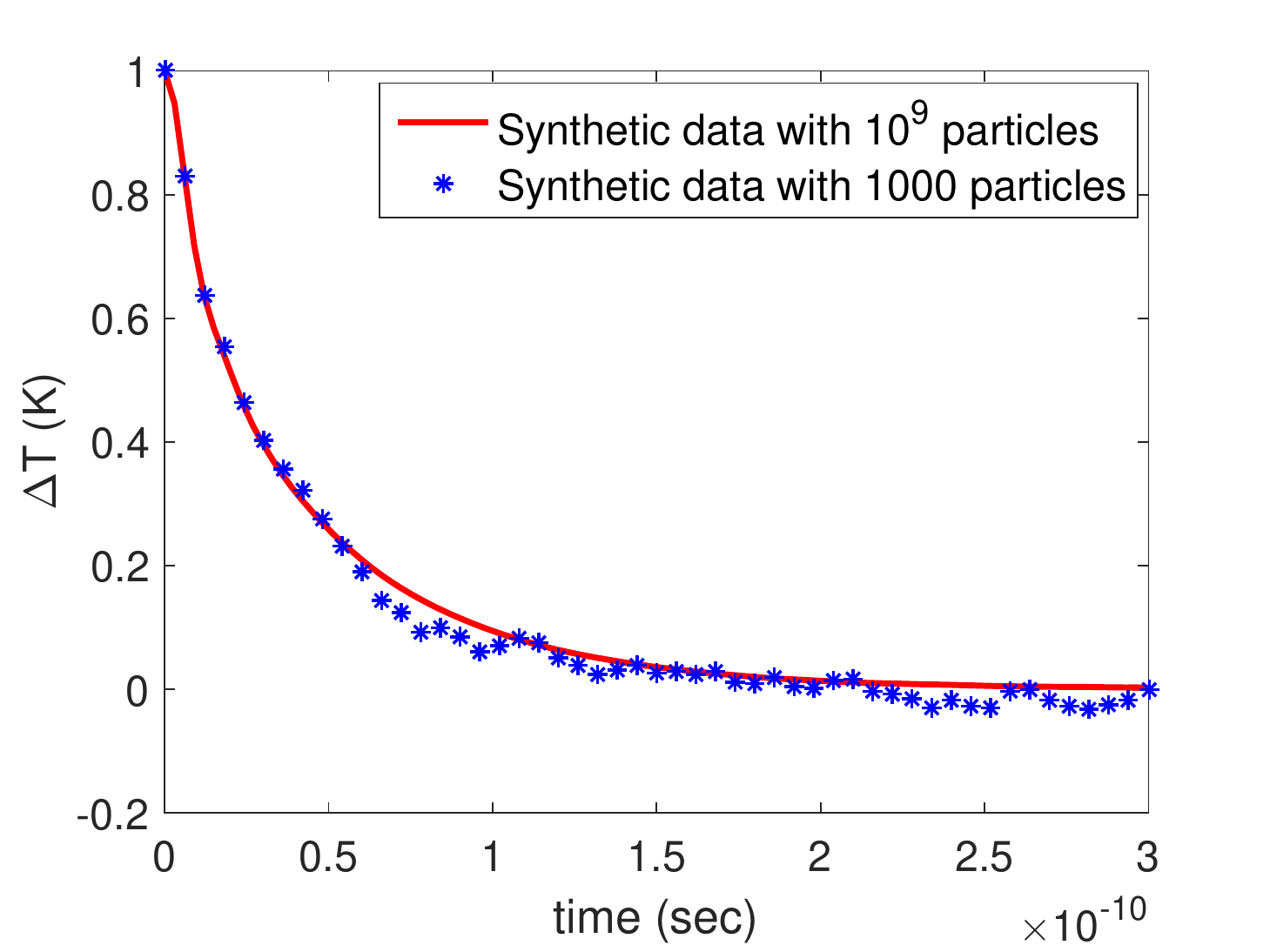}
\caption{temperature profile for $L= 100\ nm$}
\label{}
\end{subfigure}
\caption{Comparison of the synthetic temperature profile obtained by MC simulations using $\mathcal{N}_m= 10^9$ and $\mathcal{N}_m= 10^3$ computational particles based on the Holland material model. Two different grating wavelengths are shown. }
\label{noisydata}
\end{figure}

The synthetic data used to obtain the results in figures \ref{compwithanalytical} and \ref{compwithMC} were generated by MC simulations using $\mathcal{N}_m= 10^9$ computational particles. This high number of computational particles leads to a very small variance, corresponding to a very accurate ``experimental measurement". In order to evaluate the performance of the reconstruction method in the presence of noise in the experimental measurement, we have also performed reconstruction calculations with noisy synthetic data. For that purpose, we have used only $\mathcal{N}_m= 10^3$ computational particles in the MC simulations that generated the synthetic data, leading to a standard deviation of $0.02\ K$. This number of particles makes the uncertainty in $T_m$ significantly larger than the noise in common experimental data (e.g. compare the noisy temperature profile of figure \ref{noisydata} with figure 2c of \cite{Lingping_nature}). Figure \ref{noisydata} shows a comparison between the synthetic data obtained with $\mathcal{N}_m= 10^9$ and $\mathcal{N}_m= 10^3$ particles.
 
\begin{figure} [H]
\begin{subfigure}{.5\textwidth}
\centering
\includegraphics[width=1.0\linewidth]{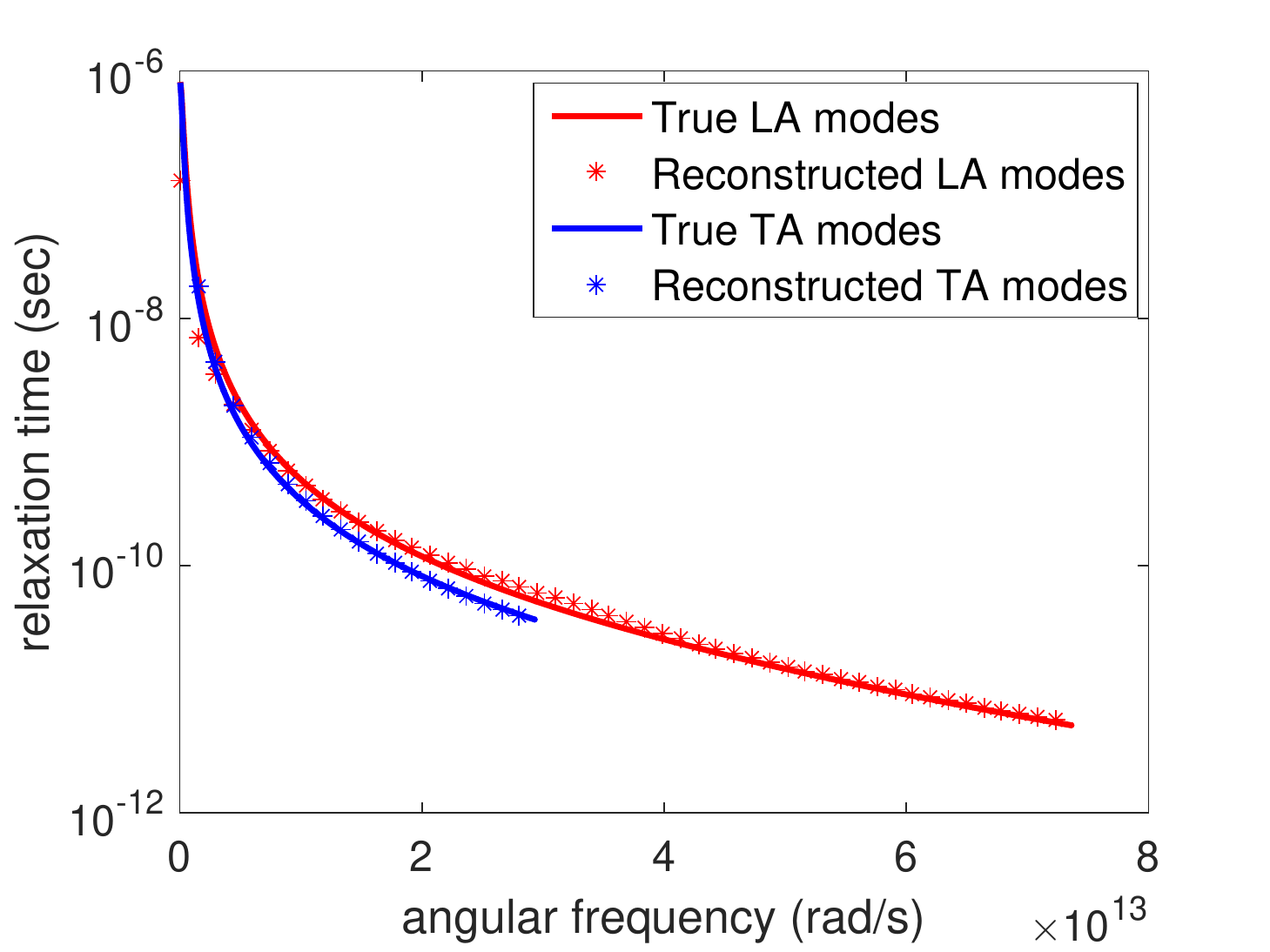}
\caption{relaxation times}
\label{}
\end{subfigure}
\begin{subfigure}{.5\textwidth}
\centering
\includegraphics[width=1.01\linewidth]{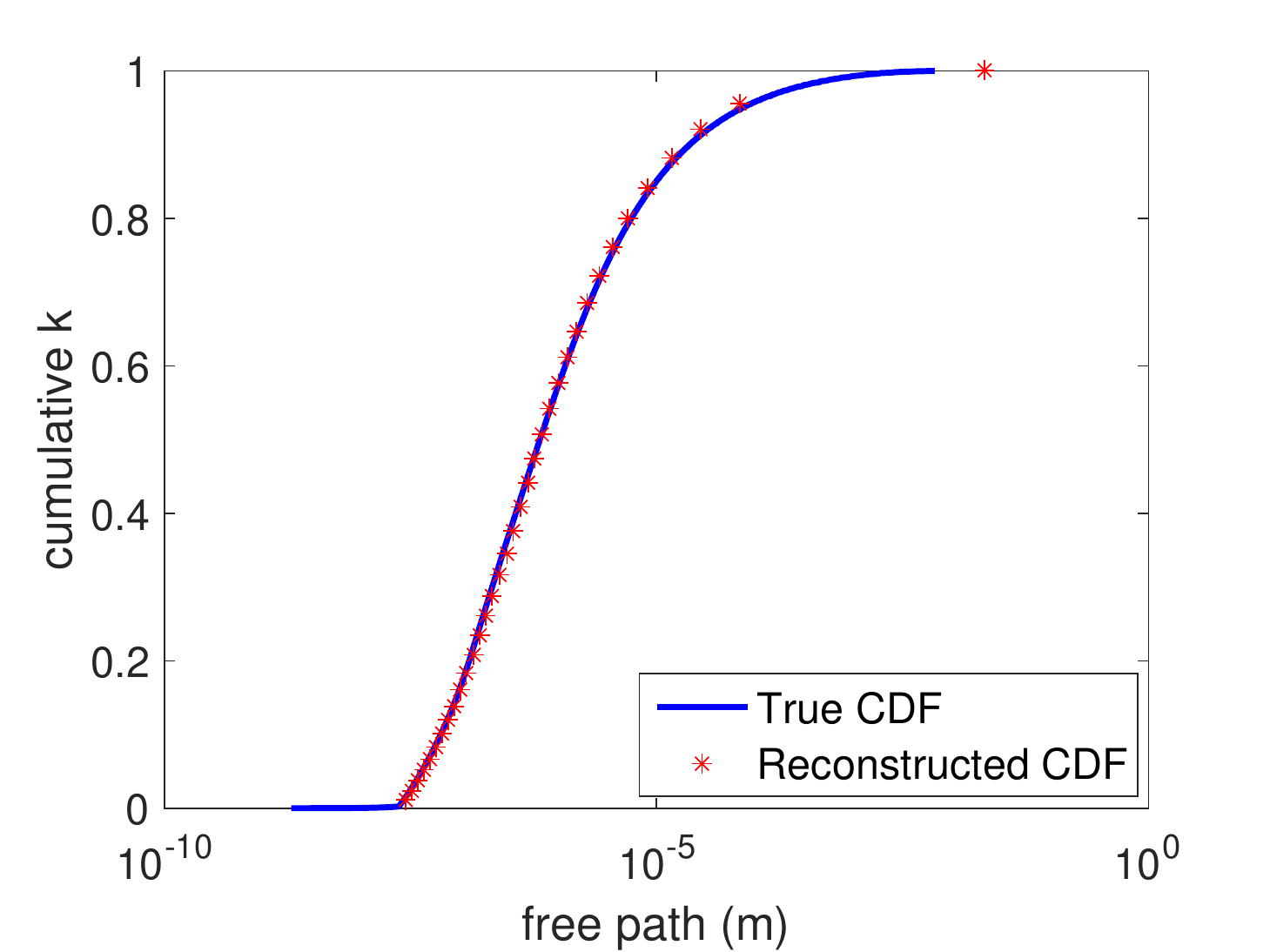}
\caption{free path distribution}
\label{}
\end{subfigure}
\caption{Comparison between true and reconstructed relaxation times and free path distribution using IFFT of \eqref{Analytical} from noisy synthetic data for the Holland material model.}
\label{noisy-analytical}
\end{figure}

\begin{figure} [H]
\begin{subfigure}{.5\textwidth}
\centering
\includegraphics[width=1.0\linewidth]{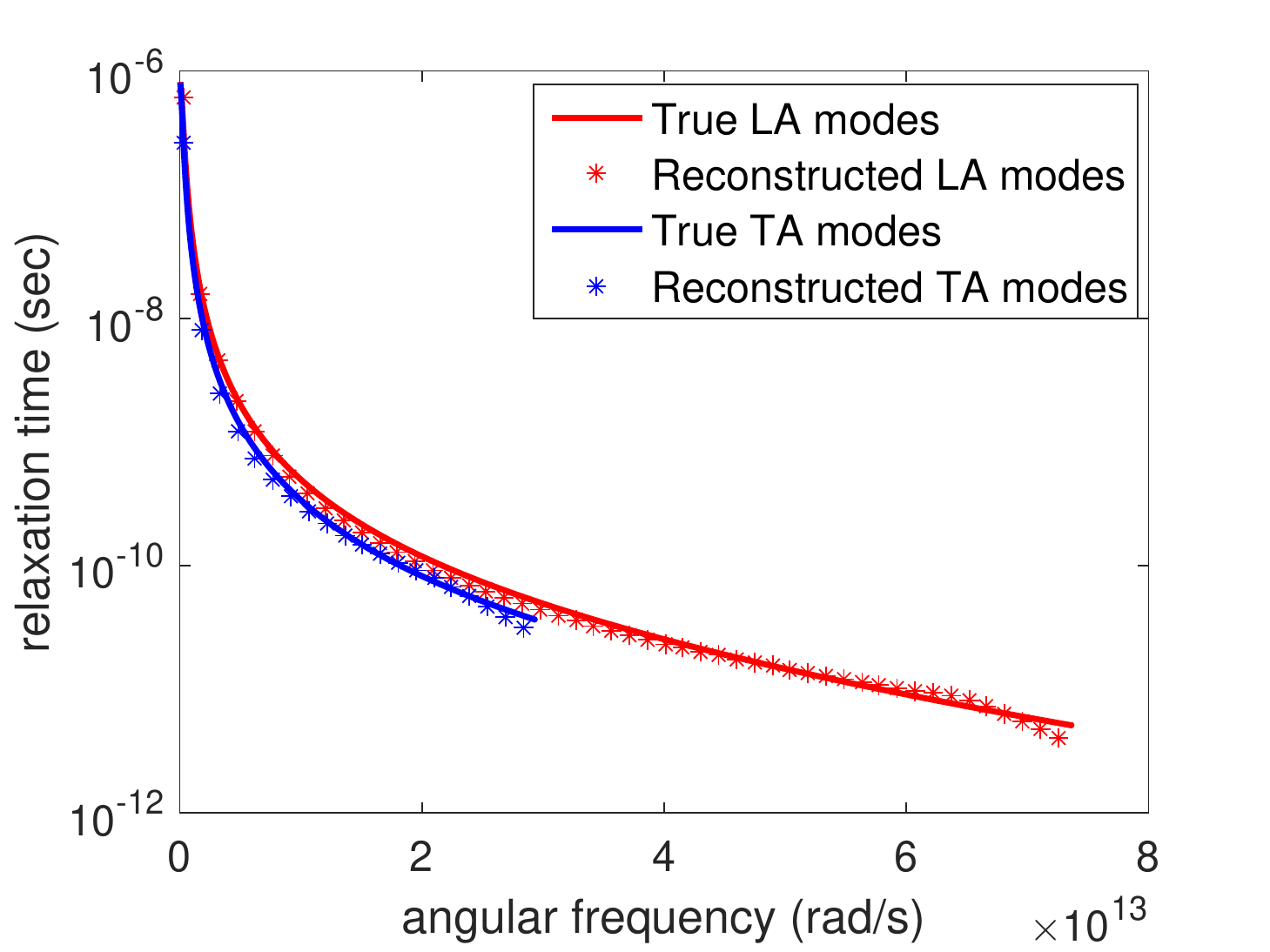}
\caption{relaxation times}
\label{}
\end{subfigure}
\begin{subfigure}{.5\textwidth}
\centering
\includegraphics[width=1.01\linewidth]{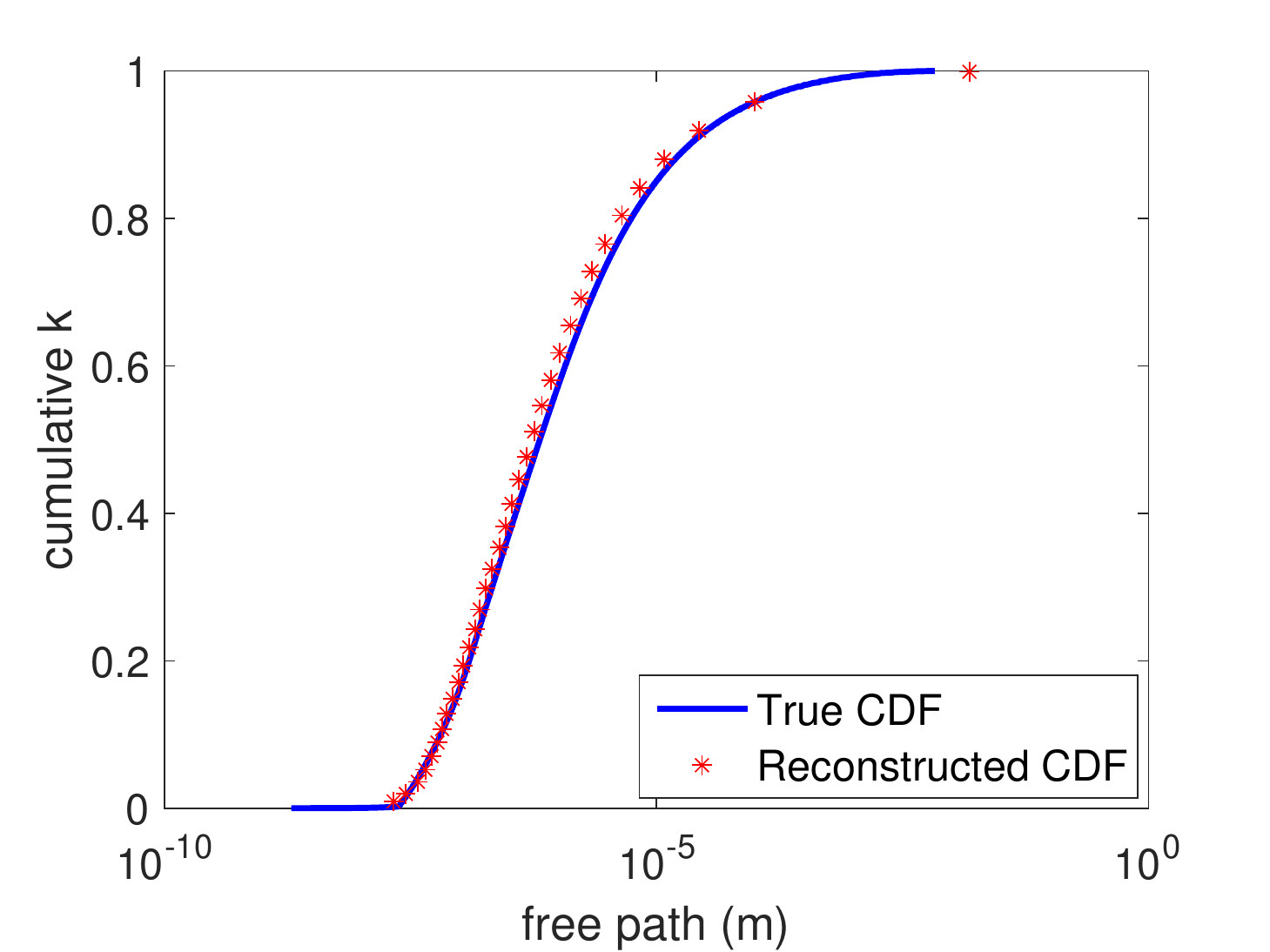}
\caption{free path distribution}
\label{}
\end{subfigure}
\caption{Comparison between true and reconstructed relaxation times and free path distribution using adjoint MC simulation from noisy synthetic data for the Holland material model.}
\label{noisy-MC}
\end{figure}

The reconstructed relaxation times and free path distribution corresponding to the noisy synthetic data ($\mathcal{N}_{m}= 10^3$) using semi-analytical and MC solutions with the Holland material model are provided in figures \ref{noisy-analytical} and \ref{noisy-MC}, respectively. In both figures we observe that even in the presence of the considerable noise in the measurement, the algorithm is able to infer the relaxation times and the free path distribution with reasonable accuracy.

\subsubsection{Ab initio model} 
\label{ab_initio}
\begin{figure} [H]
\begin{subfigure}{.5\textwidth}
\centering
\includegraphics[width=0.95\linewidth]{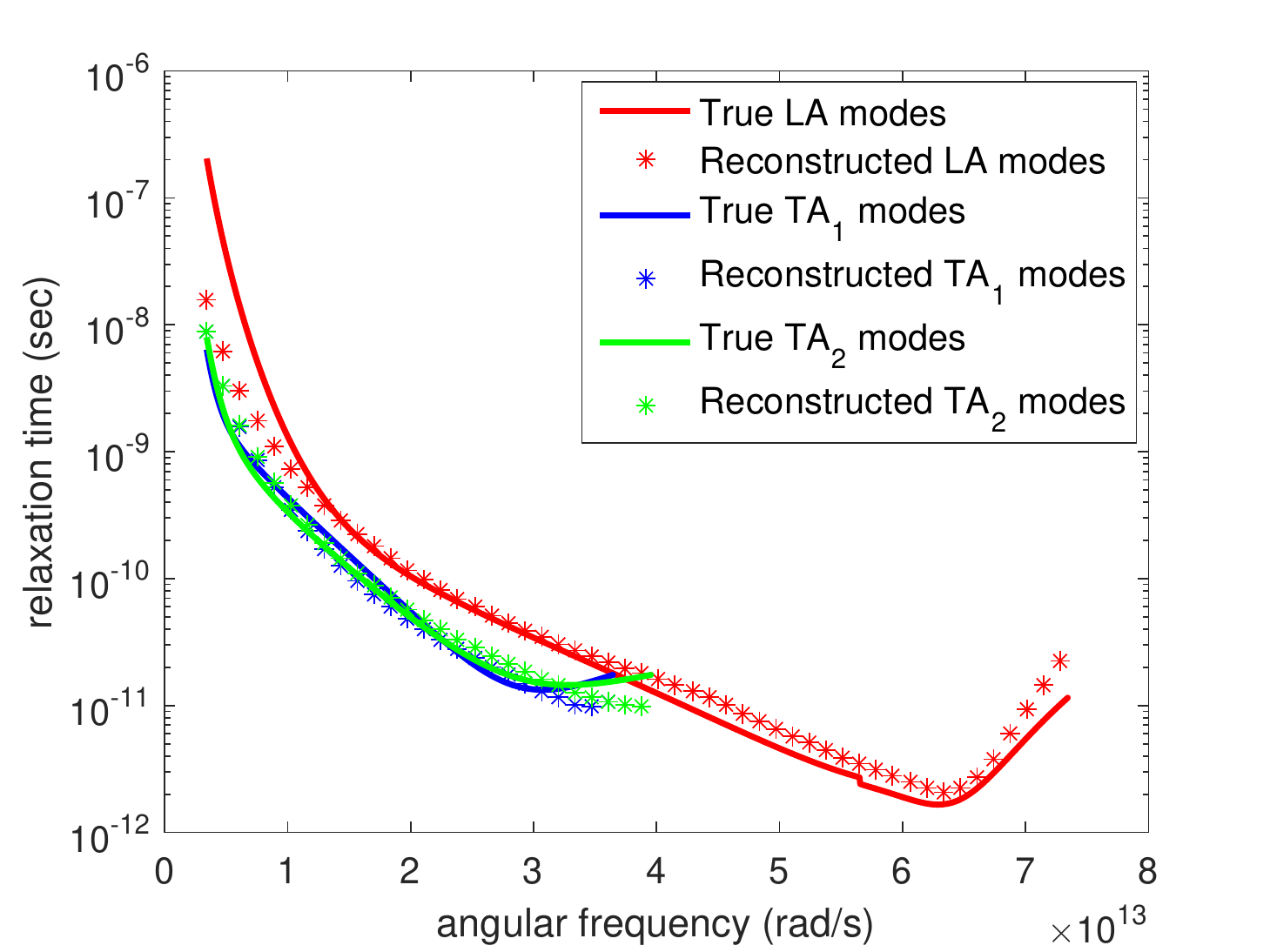}
\caption{relaxation times}
\label{}
\end{subfigure}
\begin{subfigure}{.5\textwidth}
\centering
\includegraphics[width=0.95\linewidth]{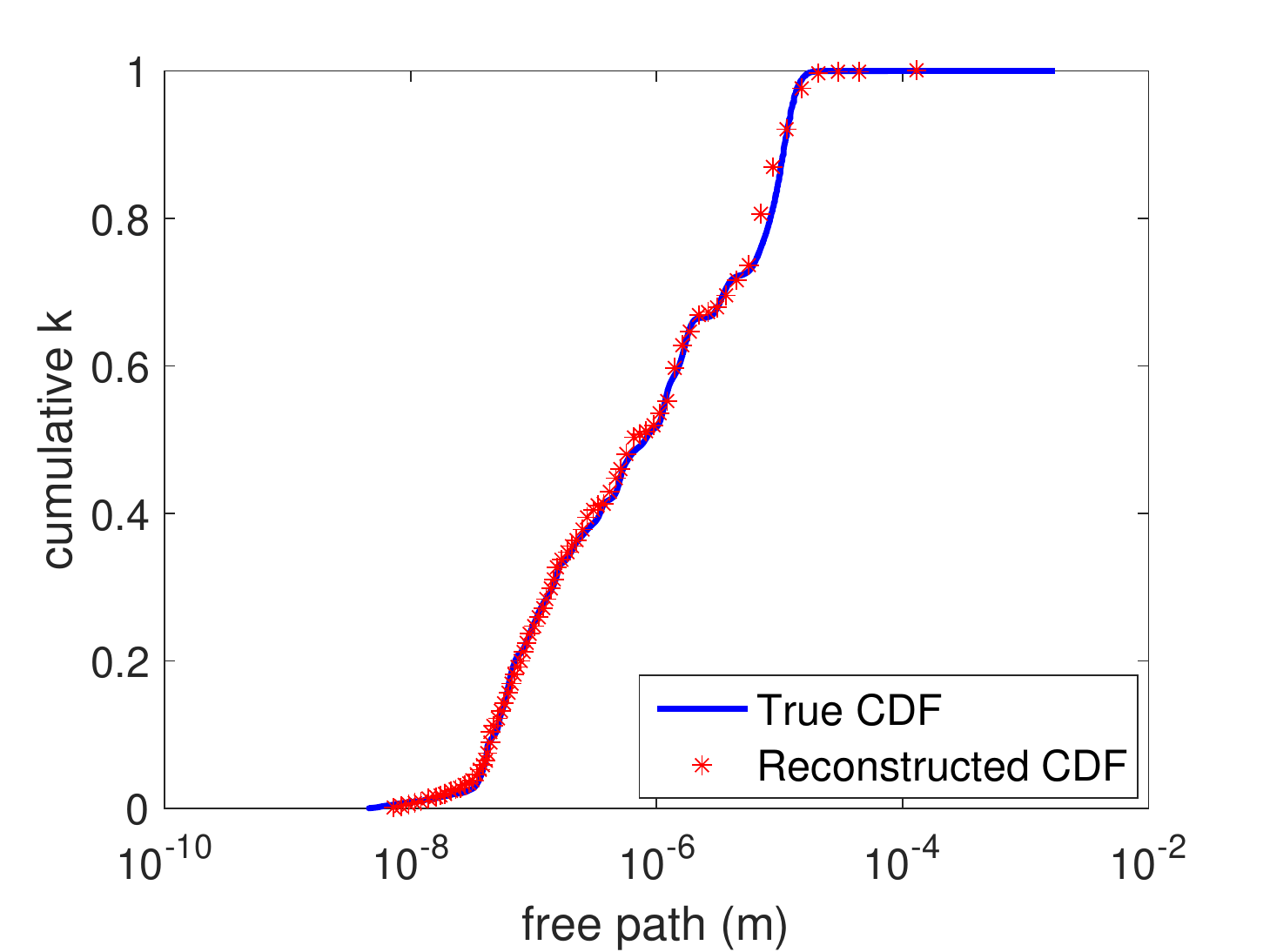}
\caption{free path distribution}
\label{}
\end{subfigure}
\caption{Comparison between true and reconstructed relaxation times and free path distribution using IFFT of \eqref{Analytical} with the ab initio material model. }
\label{compwithanalytical_2TA}
\end{figure}

\begin{figure} [H]
\begin{subfigure}{.5\textwidth}
\centering
\includegraphics[width=0.95\linewidth]{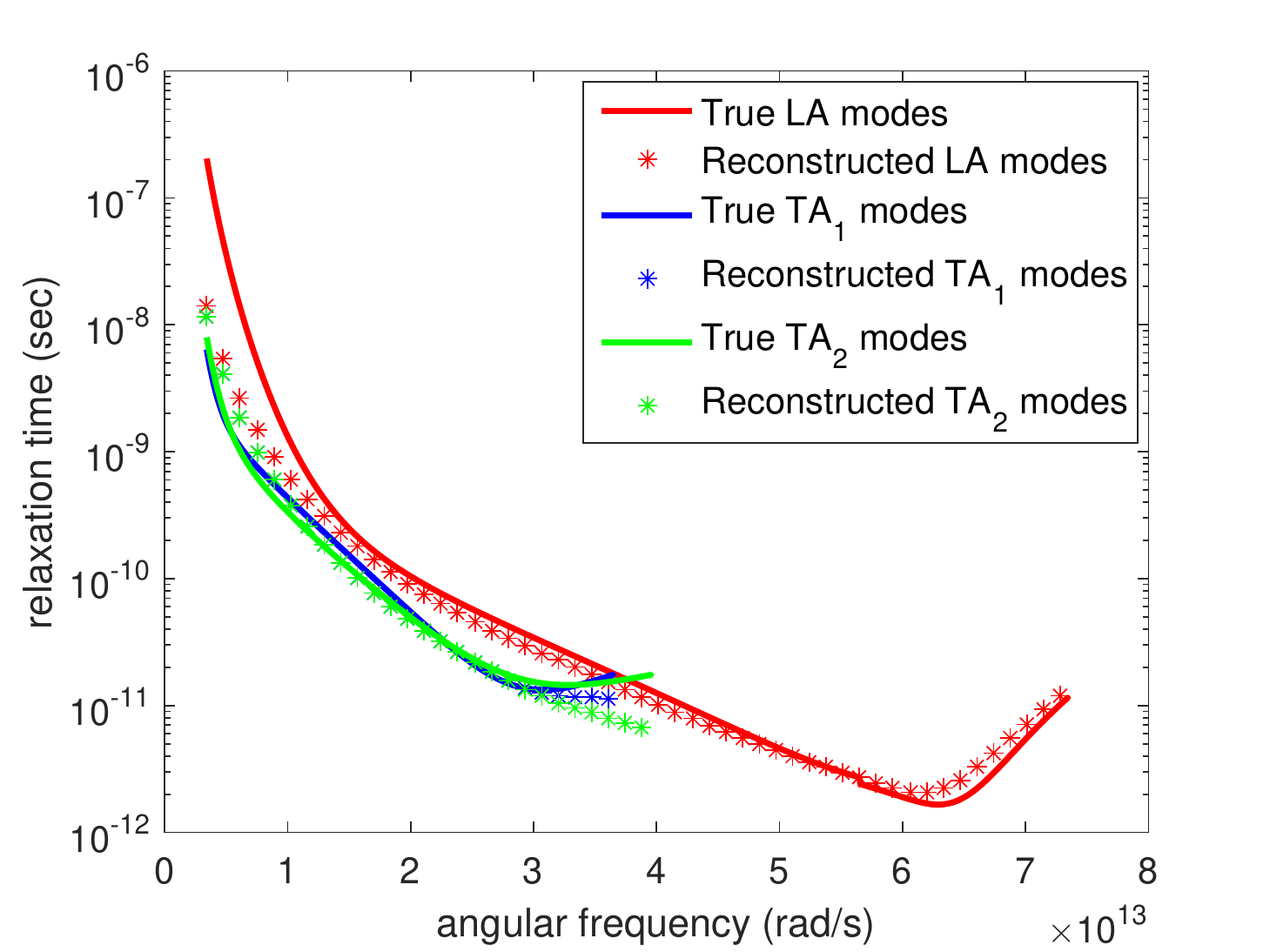}
\caption{relaxation times}
\label{}
\end{subfigure}
\begin{subfigure}{.5\textwidth}
\centering
\includegraphics[width=0.95\linewidth]{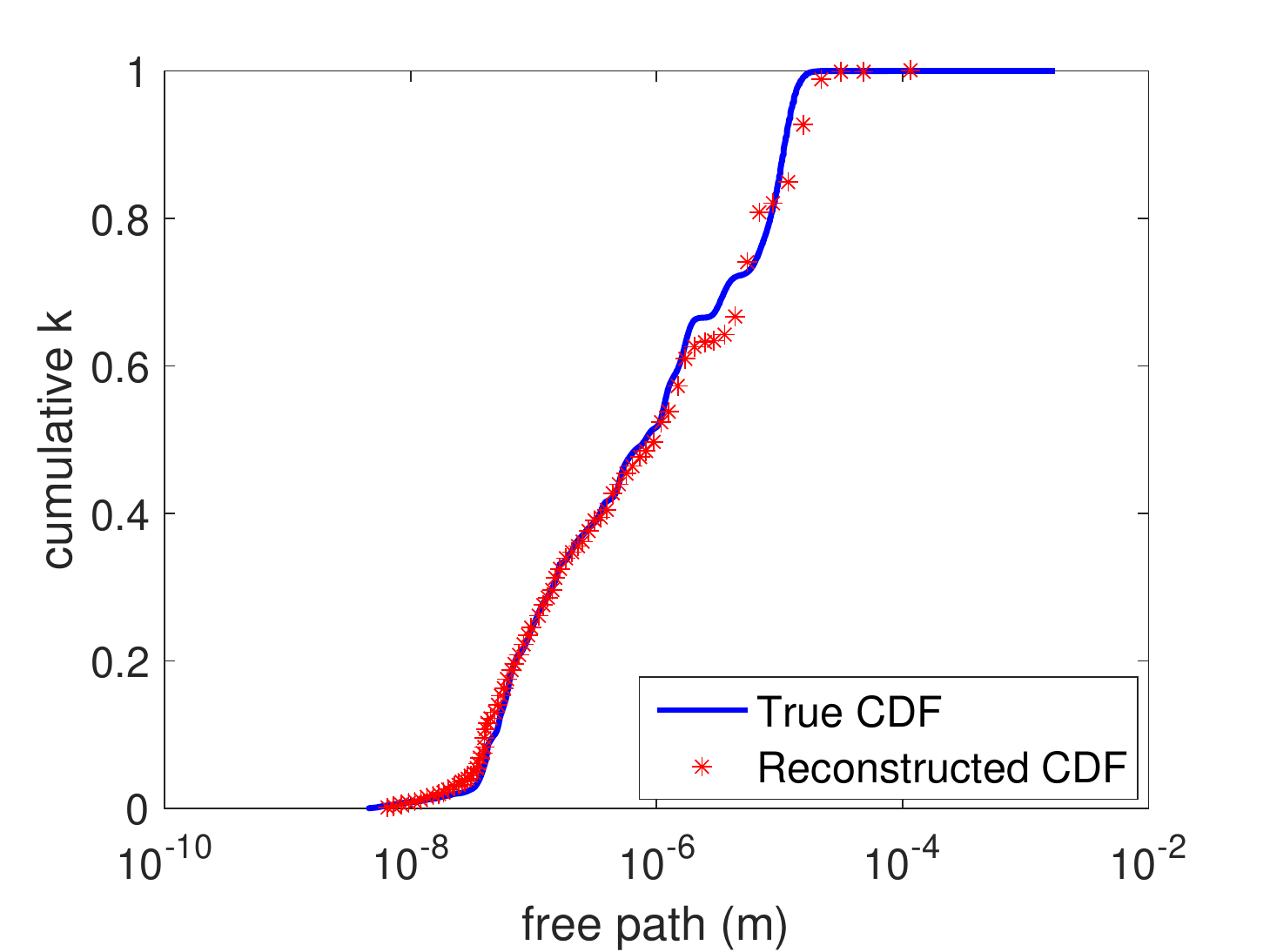}
\caption{free path distribution}
\label{}
\end{subfigure}
\caption{Comparison between true and reconstructed relaxation times and free path distribution using $T_{BTE}$ obtained from adjoint MC simulation of the ab initio material model. }
\label{compwithMC_2TA}
\end{figure}

\begin{figure} [H]
\begin{subfigure}{.5\textwidth}
\centering
\includegraphics[width=1.0\linewidth]{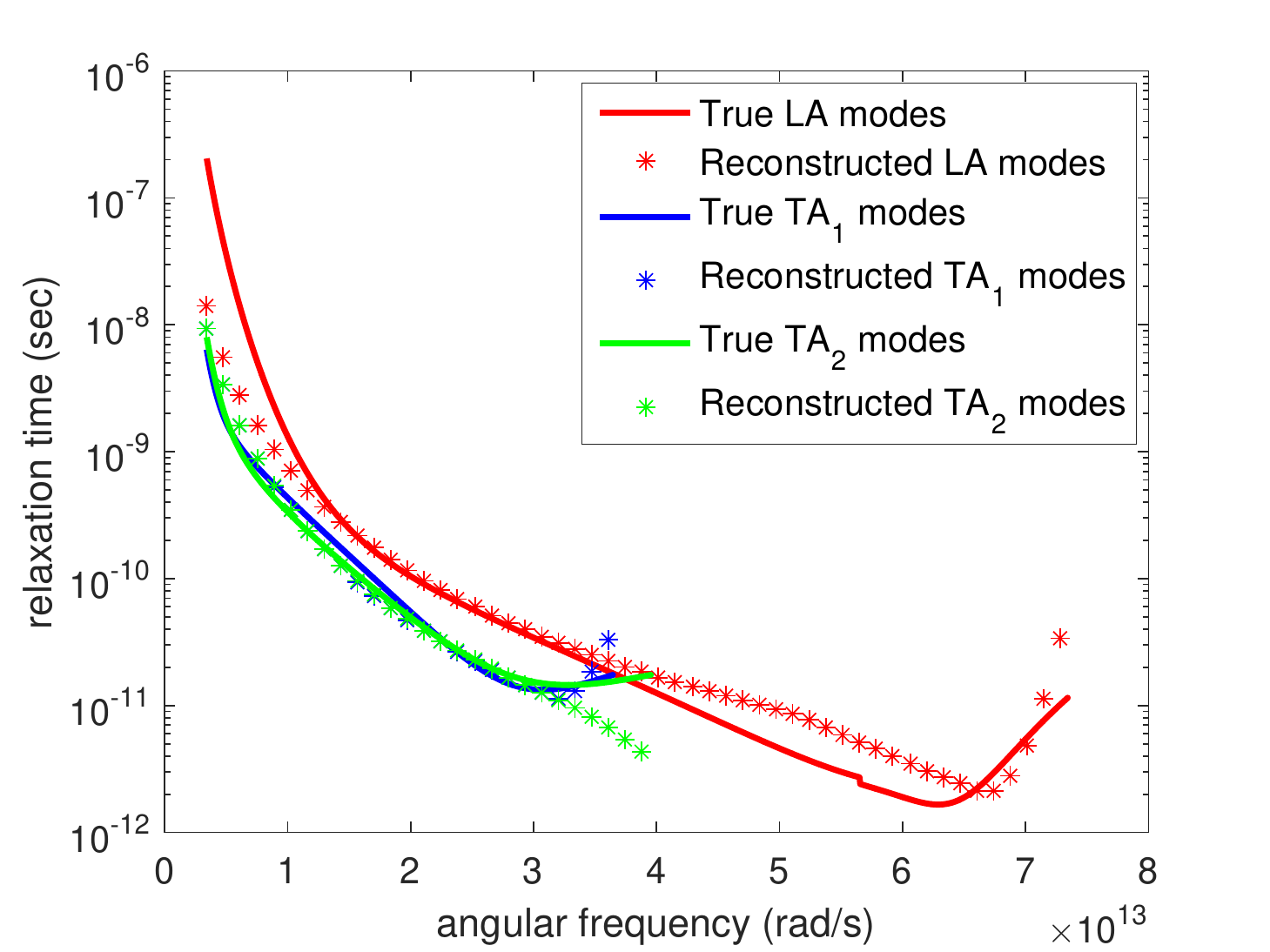}
\caption{relaxation times}
\label{}
\end{subfigure}
\begin{subfigure}{.5\textwidth}
\centering
\includegraphics[width=1.01\linewidth]{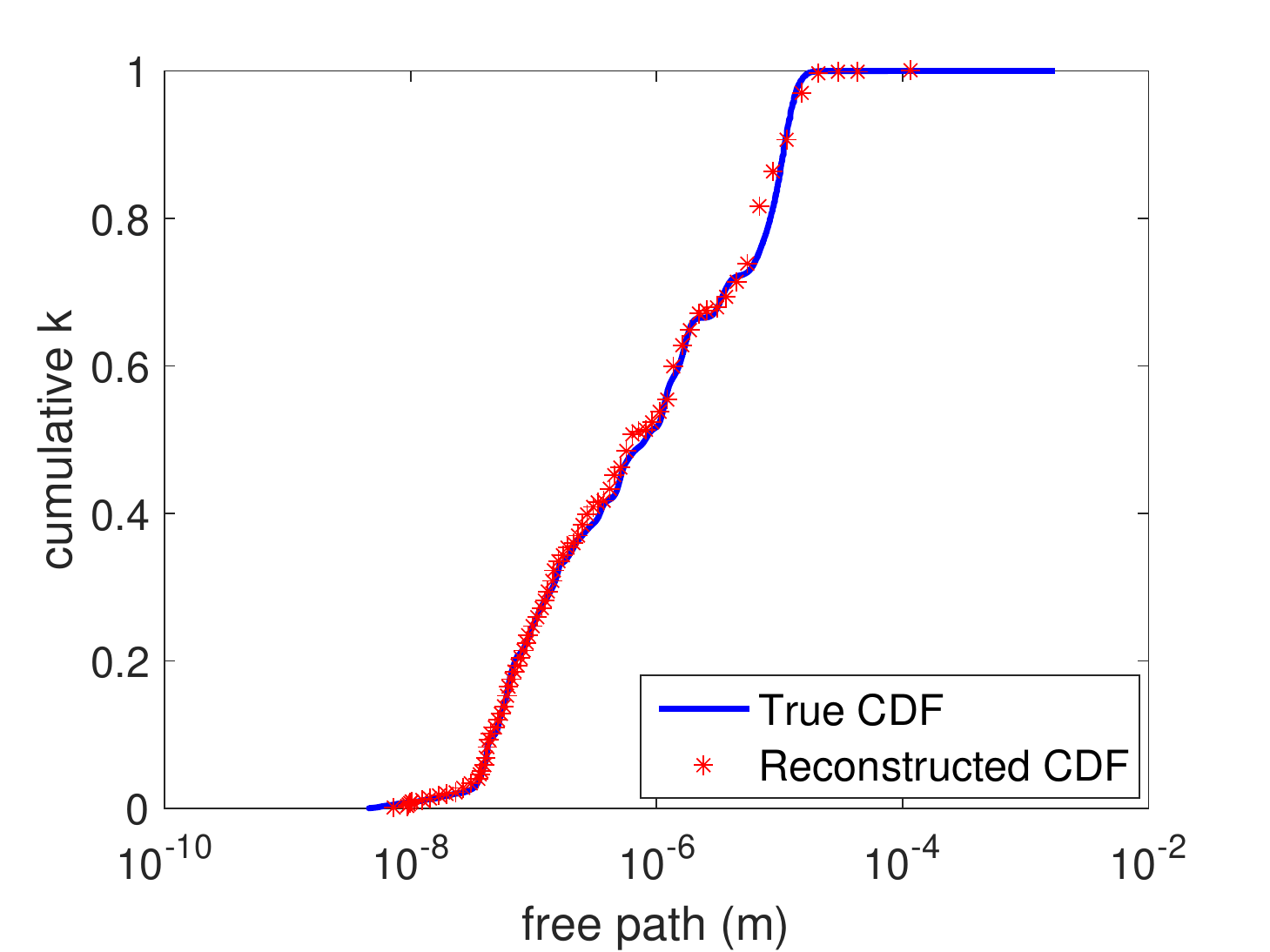}
\caption{free path distribution}
\label{}
\end{subfigure}
\caption{Comparison between true and reconstructed relaxation times and free path distribution using IFFT of \eqref{Analytical} from noisy synthetic data for the ab initio material model.}
\label{noisy-analytical_2TA}
\end{figure}

\begin{figure} [H]
\begin{subfigure}{.5\textwidth}
\centering
\includegraphics[width=1.0\linewidth]{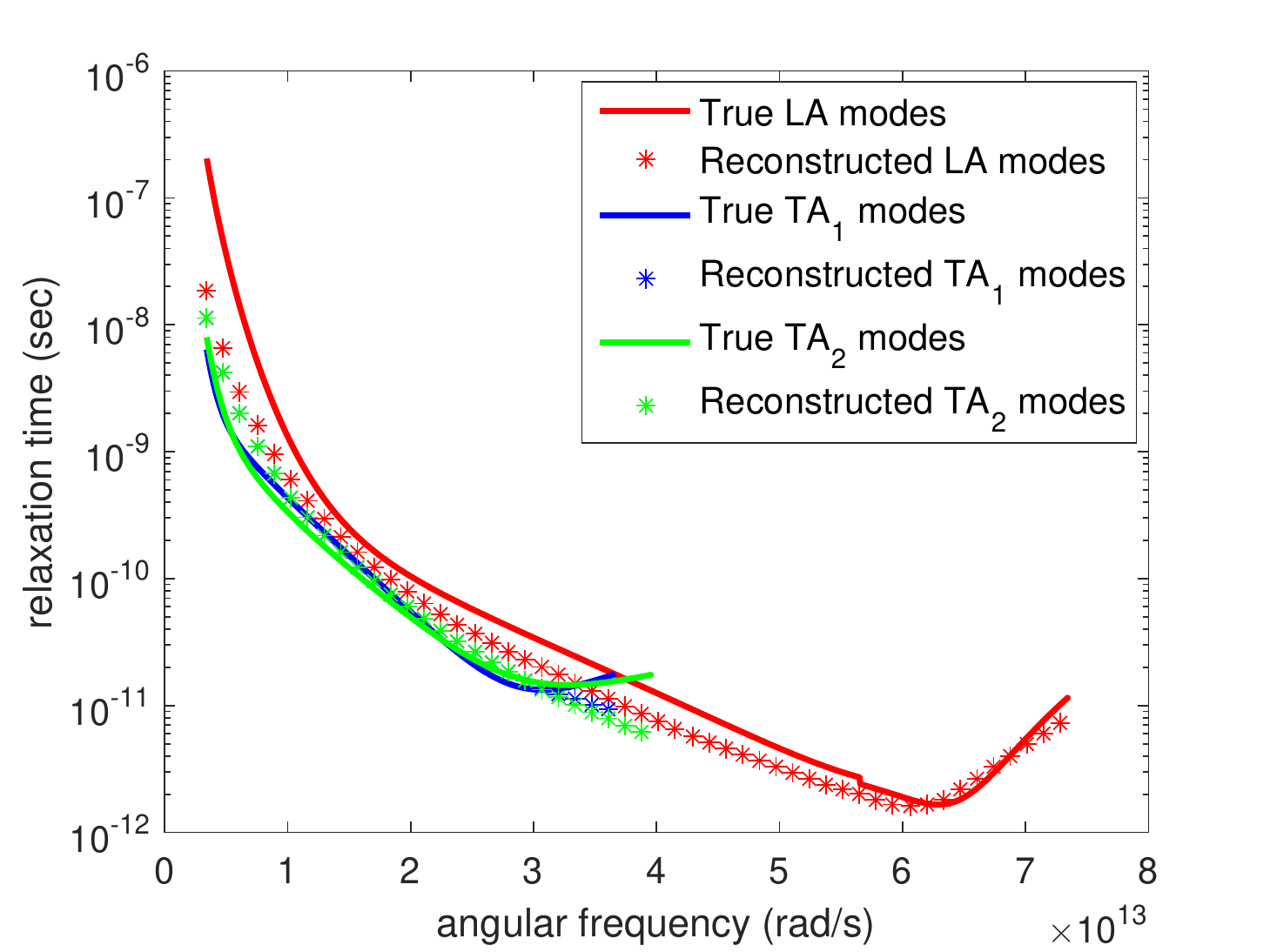}
\caption{relaxation times}
\label{}
\end{subfigure}
\begin{subfigure}{.5\textwidth}
\centering
\includegraphics[width=1.01\linewidth]{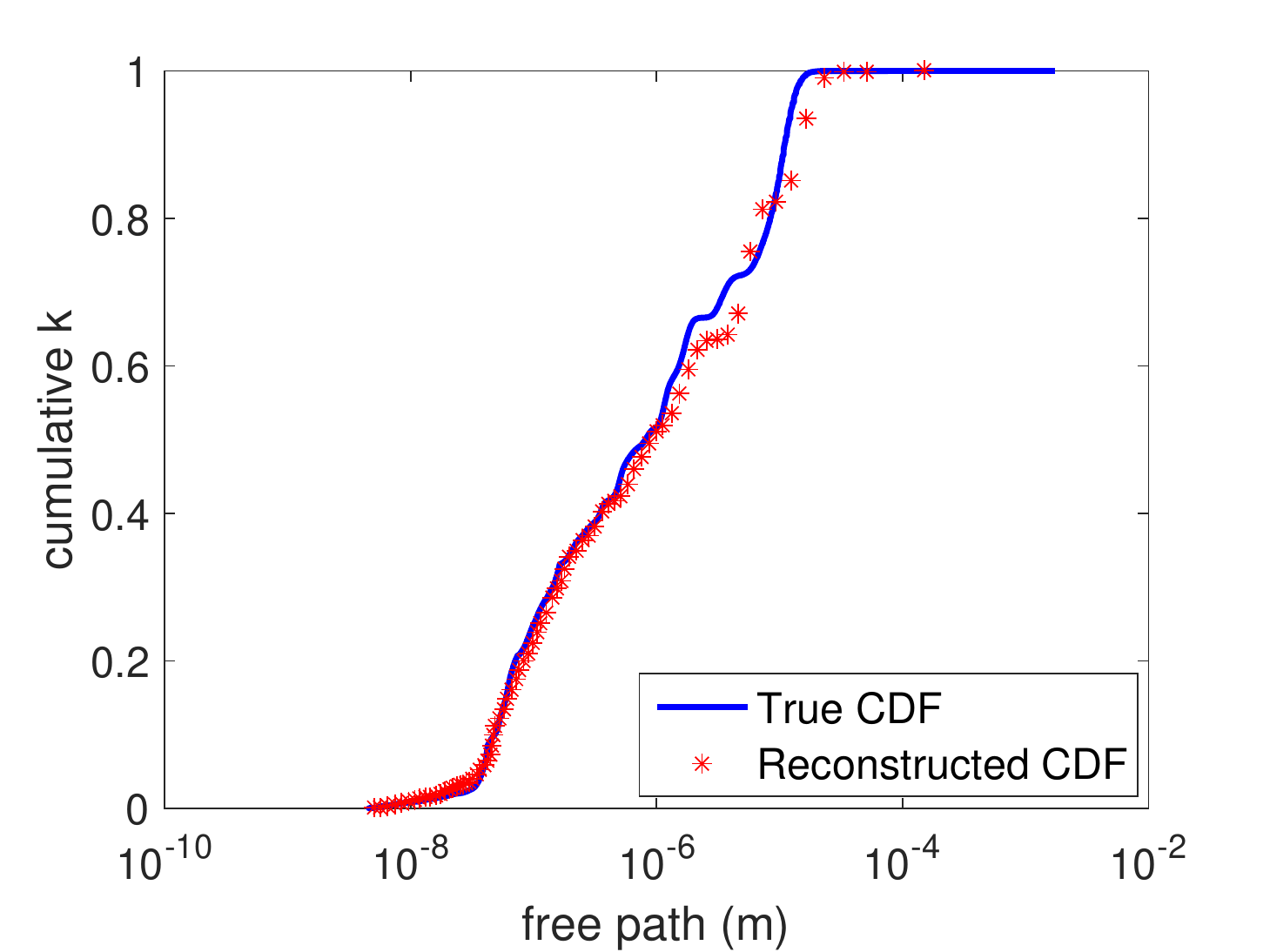}
\caption{free path distribution}
\label{}
\end{subfigure}
\caption{Comparison between true and reconstructed relaxation times and free path distribution using adjoint MC simulation from noisy synthetic data for the ab initio material model.}
\label{noisy-MC_2TA}
\end{figure}

In this section, we repeat the validation process, namely using IFFT of \eqref{Analytical} and adjoint MC simulation in the presence and absence of noise in the synthetic data, for the ab initio material model. The reconstructed material properties in the absence of noise ($\mathcal{N}_{m}= 10^9$) are provided in figures \ref{compwithanalytical_2TA} and \ref{compwithMC_2TA}. The reconstructed material properties in the presence of noise ($\mathcal{N}_{m}= 10^3$) using IFFT of \eqref{Analytical}  and MC simulation are provided in figures \ref{noisy-analytical_2TA} and \ref{noisy-MC_2TA}, respectively.

We observe that the general level of agreement is very good, but not at the level of the Holland model discussed in section \ref{Holland}. This is clearly to be expected given the relative complexity of the two models and the fact that the parametrization \eqref{parametrization_2TA} is related to the Holland model in limiting cases. We also note that although the discrepancy is particularly noticeable for low frequencies, this is primarily an artifact of the low density of states of modes in these frequency ranges. Specifically, the density of states of the $LA$ modes for $\omega \leq 8 \times 10^{12}\ rad/s$ is zero (i.e. none of the two terms in \eqref{objectivefunction_2TA} influences $\tau^{LA}_{\omega}$ for $\omega \leq 8 \times 10^{12} )$). The role of density of states can be further verified by noting that the reconstructed free path distribution is in all cases in better agreement with the input (``true") data compared to the corresponding relaxation times. 

We also observe that, as in the case of the Holland model, the noise in the synthetic data has only a small effect on the reconstruction quality.

\section{Comparison to effective thermal conductivity approach} \label{discussion}
\begin{figure} [H]
\begin{subfigure}{.5\textwidth}
\centering
\includegraphics[width=1.0\linewidth]{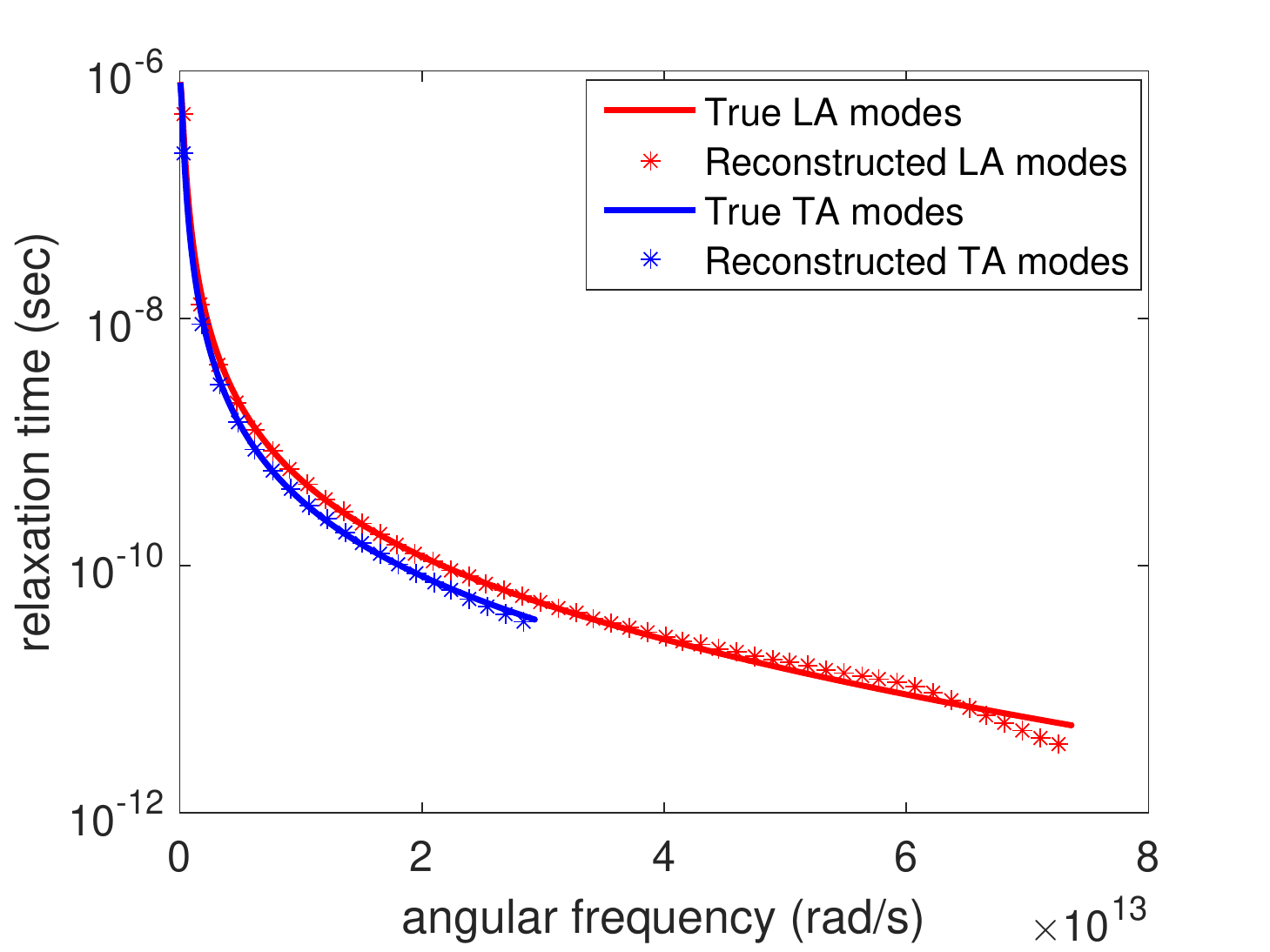}
\caption{relaxation times}
\label{}
\end{subfigure}
\begin{subfigure}{.5\textwidth}
\centering
\includegraphics[width=1.01\linewidth]{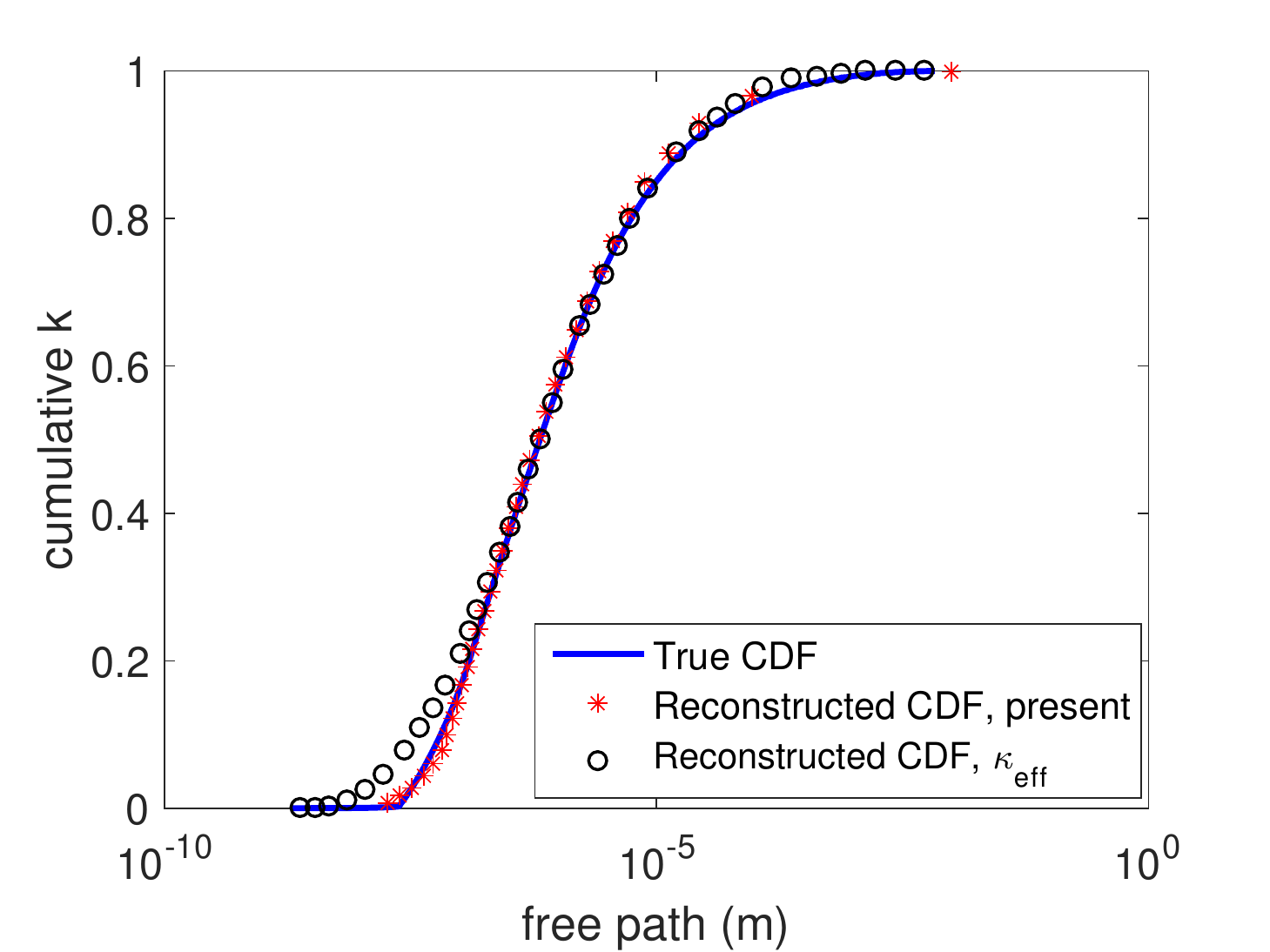}
\caption{free path distribution}
\label{}
\end{subfigure}
\caption{Comparison between true and reconstructed relaxation times and free path distribution using IFFT of the semi-analytical result \eqref{Analytical} for the Holland material model with limited range of input wavelengths ($L= 1\ \mu m, 5\ \mu m, 10\ \mu m,$ and $50\ \mu m$). Relaxation data was ``recorded" in the time period $0 \leq t \leq t_L$, where $t_L= \min (t_{L, 1\%}, 50\ ns)$. The ``Reconstructed CDF, $\kappa_{eff}$" curve corresponds to data transcribed \cite{explanation} from figure 2 of \cite{Minnich2012}, which uses the effective thermal conductivity approach with 12 relaxation profiles with wavelengths in the range $L= 1-15\ \mu m$. }
\label{lessdata}
\end{figure}

The accuracy of the reconstruction process shown in figures \ref{compwithanalytical}, \ref{compwithMC}, \ref{noisy-analytical}-\ref{noisy-MC_2TA} is very encouraging and, generally speaking, superior to that obtained by effective-thermal-conductivity-based approaches (see, for example, figure 2 in \cite{Minnich2012}, or figure \ref{lessdata} of present manuscript). In \cite{Minnich2012}, it is pointed out that the effective-thermal-conductivity-based approach is less accurate outside the range of lengthscales over which input data is available and thus the discrepancies observed in \cite{Minnich2012} may be attributed to the lack of data in the range $L \lesssim 1\ \mu m$ and $L \gtrsim 15\ \mu m$. On the other hand, as shown in figure \ref{lessdata},  reconstruction using the method proposed here with $L= 1\ \mu m, 5\ \mu m, 10\ \mu m,$ and $50\ \mu m$, reveals that this method is minimally  sensitive to the range in which data exist.  

Sensitivity to the range over which input data exists is very important, not only because some lengthscales may not be experimentally accessible \cite{Minnich2012}, but also because  the effective thermal conductivity concept cannot be used to analyze small lengthscale or early-time data---recall that Fourier-based descriptions are only accurate in the limit of large lengthscales (compared to the mean free path) and long timescales (compared to the mean time between scattering events). Below, we theoretically analyze the limitations arising from these  requirements by studying in detail the response associated with the TTG setup.

Another problem associated with the effective thermal conductivity approach is that, other than in the thermal grating geometry, the associated suppression functions are unknown (and depend not only on the experimental setup, but also on the material parameters that are being calculated). Using gray MC simulations as a substitute for the suppression function introduces approximations (see \cite{Collins2013} for a discussion) that are not always acceptable and more importantly not {\it a priori} controllable.

\subsection{The thermal grating case in detail} 
\label{TTG_discussion}
Clearly, in order to extract information from the experiment correctly, a model needs to be able to fit the functional form of the experimental response accurately. We now examine, both theoretically and via simulation, if this requirement is satisfied by the effective thermal conductivity formalism in the case of the TTG problem.

We first recall that the effective thermal conductivity concept applies to late times, as defined by $\tau_{\omega} \zeta \ll 1$ \cite{Hua2014} ($\zeta$ is the Fourier transform variable with respect to time, see discussion in section \ref{TTG}). Following previous work \cite{Hua2014}, we simplify \eqref{S_omega} in the limit of  $\tau_{\omega} \zeta \ll 1$  to obtain (see Appendix C)
\begin{equation}
S_{\omega}= \frac{C_{\omega}} { \tau_{\omega} } \left[ \frac{\tan^{-1} \left( K\!n_{\omega} \right) } {K\!n_{\omega}}- \frac{\tau_{\omega} } {K\!n_{\omega}^2+ 1 } i \zeta \right] .
\label{S_omega_quasi}
\end{equation}
Here, $K\!n_{\omega}= 2 \pi v_{\omega} \tau_{\omega}/L$.
Substituting equation \eqref{S_omega_quasi} in \eqref{Analytical} and performing the inverse Fourier transform in time leads to the following temperature response in the {\it late time limit} $t \gg \tau_{\omega}$
\begin{equation}
\Delta T= A \exp \left( \frac{2 \pi i x} {L} \right) \exp \left(-\left( \frac{2 \pi} {L} \right)^2 \frac{\kappa_{mod}} {C_{mod}} t \right) ,
\label{T_quasi}
\end{equation}
where 
\begin{equation}
\kappa_{mod}= \int_{\omega} \frac{1} {3} C_{\omega} v_{\omega}^2 \tau_{\omega} \left\{ \frac{3} {K\!n_{\omega}^2} \left[ 1- \frac{\tan^{-1} \left( K\!n_{\omega} \right)} {K\!n_{\omega}} \right] \right\} d \omega 
\label{k_quasi}
\end{equation}
is the effective thermal conductivity as found before \cite{Hua2014} and
\begin{equation}
A= \frac{ \left[ \int_{\omega} \frac{C_{\omega} \tau_{\omega}}{K\!n_{\omega}^2+ 1} d \omega \int_{\omega} \frac{C_{\omega}} {\tau_{\omega}} \left( 1- \frac{\tan^{-1} \left( K\!n_{\omega} \right)} {K\!n_{\omega}} \right) d \omega+ \int_{\omega} \frac{C_{\omega}} {K\!n_{\omega}^2+ 1} d \omega \int_{\omega} \frac{C_{\omega}} {K\!n_{\omega}} \tan^{-1} \left( K\!n_{\omega} \right) d \omega \right]^2 } { \left[ \int_{\omega} \frac{C_{\omega}} {K\!n_{\omega}^2+ 1} d \omega \right]^3}
\label{A}
\end{equation}
is the response amplitude, and
\begin{equation}
C_{mod}= \int_{\omega} \frac{C_{\omega} } {K\!n_{\omega}^2+ 1} d \omega
\label{C_eff}
\end{equation}
is an effective heat capacity. 

In other words, although the response \eqref{T_quasi} is of exponential form, it is not a solution of the Fourier heat equation with an effective thermal conductivity $\kappa_{mod}$, as usually assumed. Response \eqref{T_quasi} is a diffusive type of solution which differs from the classical Fourier solution by featuring
\begin{itemize}
\item an amplitude, $A$, different from the original temperature perturbation in the TTG experiment (taken to be $1\ K$ here)
\item a thermal conductivity given by $\kappa_{mod}$
\item a heat capacity given by $C_{mod}$
\end{itemize}
Moreover, we always need to keep in mind that the above is only true for $\tau_{\omega} \zeta \ll 1$, which turns out to have significant implications. Using the fact that $\kappa_{mod}/C_{mod} \sim O(\kappa/C) \sim \langle \Lambda_{\omega} \rangle^2 / \langle \tau_{\omega} \rangle$ we can write \eqref{T_quasi} in the form 
\begin{equation}
\Delta T \sim A \exp \left( \frac{2 \pi i x}{L} \right) \exp \left (- \pi^2 K\!n^2 \frac{t} {\langle \tau_{\omega} \rangle} \right) ,
\label{T_quasi2}
\end{equation}
where $K\!n= \langle \Lambda_{\omega} \rangle /L$ is the (average) Knudsen number,  $\langle \Lambda_{\omega} \rangle$ is the mean free path and $\langle \tau_{\omega} \rangle$ is the mean free time.
This leads us to conclude that the requirement $t/ \tau_{\omega} \gg 1$  ($t/ \langle \tau_{\omega} \rangle \gg 1$) is incompatible with a solution of this form ($\Delta T \rightarrow 0$) unless $K\!n^2 \ll 1$. This ``mathematical" incompatibility is a simple statement of the physical fact that relaxation times scale with the system size and thus for a small system not satisfying $K\!n^2 \ll  1$ the response timescale will be on the order of the relaxation time or smaller (no experimentally measurable signal will be available at times much longer than the relaxation time)---in other words, no fully diffusive relaxation regime is possible for systems not satisfying $K\!n^2 \ll 1$.  

\begin{figure} [H]
\begin{subfigure}{.5\textwidth}
\centering
\includegraphics[width=1.0\linewidth]{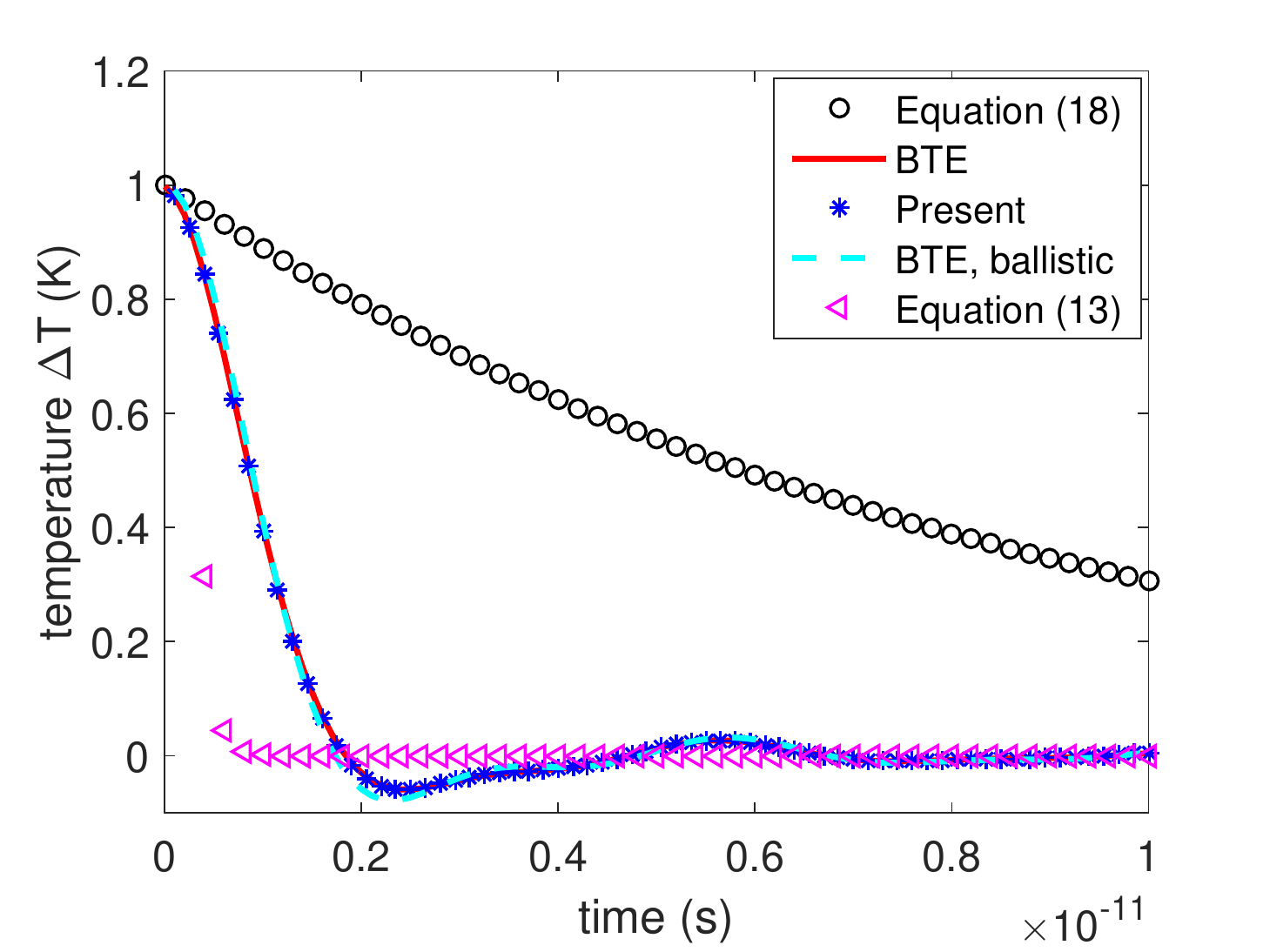}
\caption{Relaxation profile for $L= 10\ nm$  }
\end{subfigure}
\begin{subfigure}{.5\textwidth}
\centering
\includegraphics[width=1.0\linewidth]{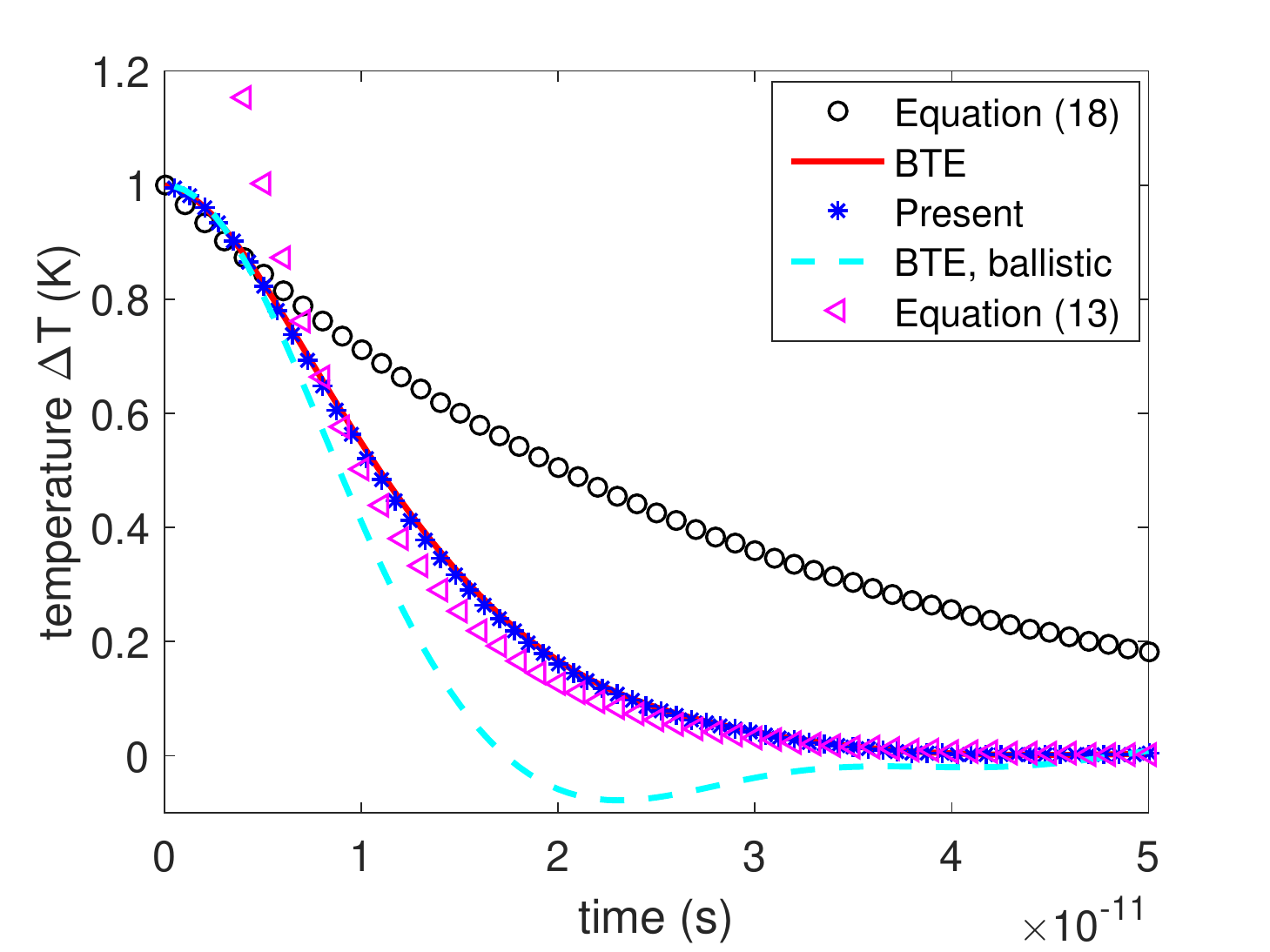}
\caption{Relaxation profile for $L= 100\ nm$ }
\end{subfigure}
\begin{subfigure}{.5\textwidth}
\centering
\includegraphics[width=1.0\linewidth]{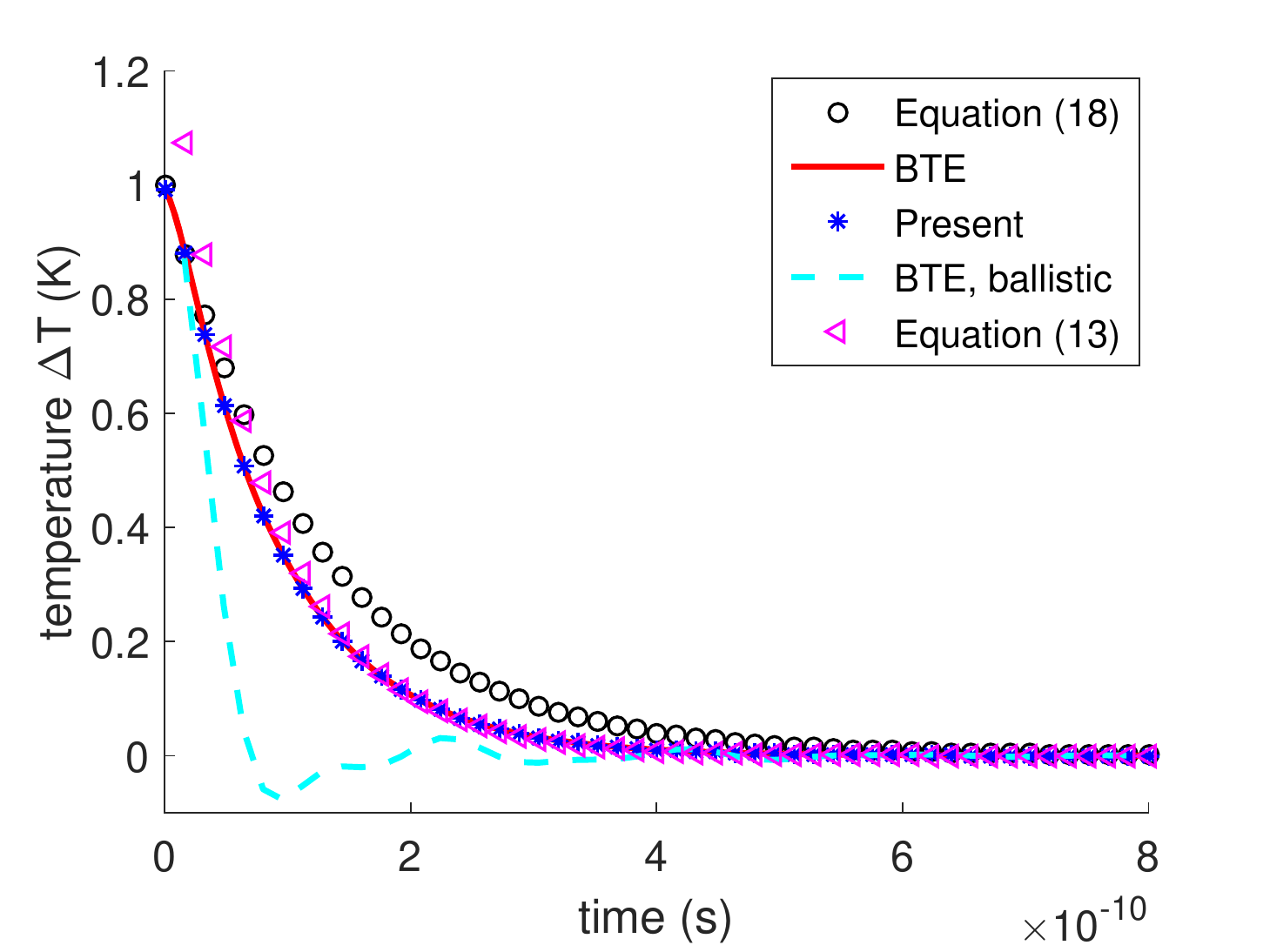}
\caption{Relaxation profile for $L= 400\ nm$ }
\label{}
\end{subfigure}
\begin{subfigure}{.5\textwidth}
\centering
\includegraphics[width=1.0\linewidth]{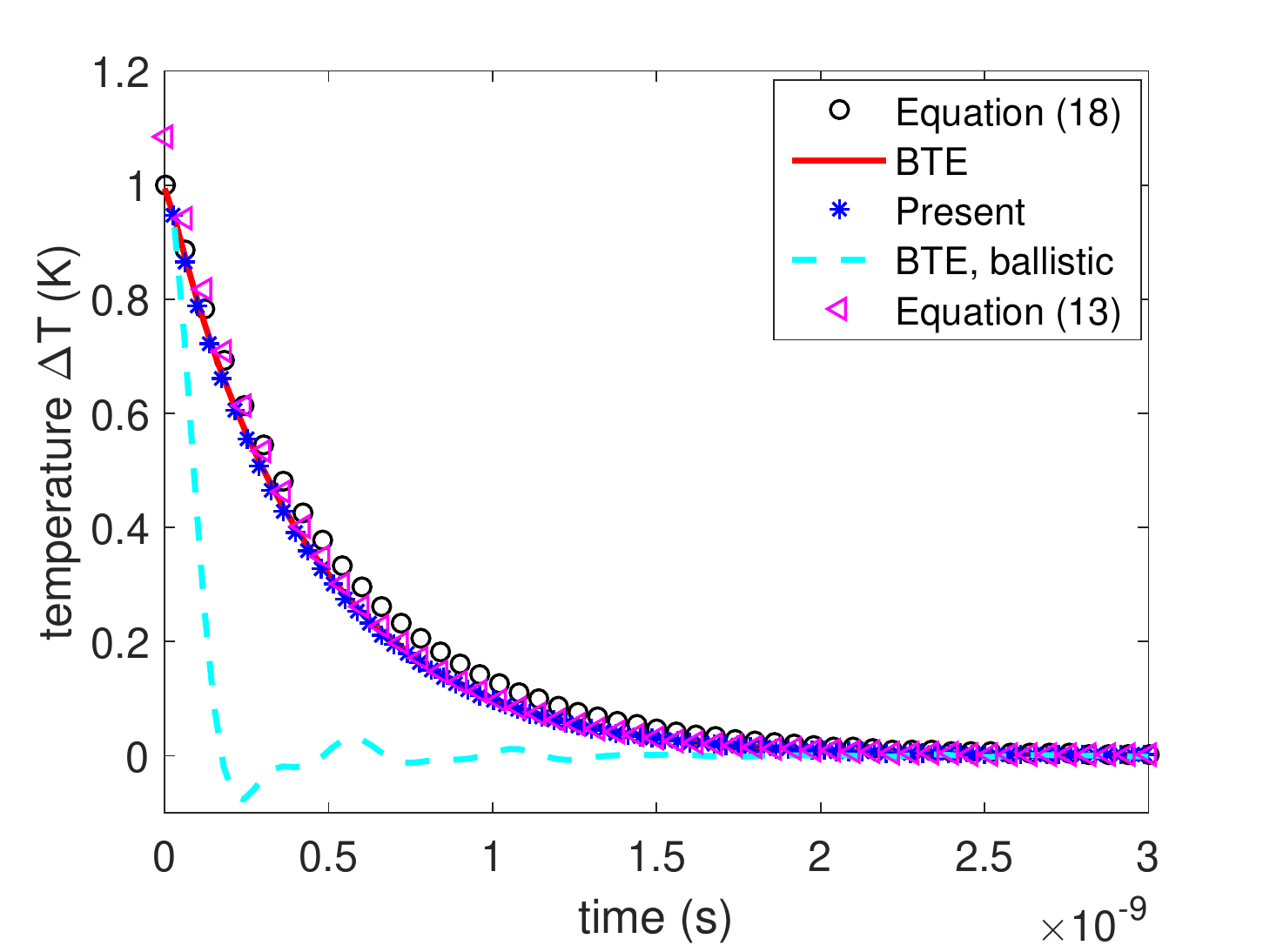}
\caption{Relaxation profile for  $L= 1\ \mu m$ }
\label{10n_T_comparison}
\end{subfigure}
\caption{Comparison between the method proposed in the present paper and various Fourier-based methods. Temperature relaxation profiles are shown for four grating wavelengths and compared to BTE solution for the ab initio material model (denoted ``BTE"). The prediction of the method proposed here, denoted as ``present", is given by solution of the Boltzmann equation based on the reconstructed relaxation time data of figure \ref{compwithanalytical_2TA}. Solutions based on equations \eqref{T_quasi} and \eqref{T_quasi_k_eff} are denoted ``Equation \eqref{T_quasi}" and ``Equation \eqref{T_quasi_k_eff}", respectively.} 
\label{Temp_comparison_2TA}
\end{figure}

\begin{figure} [H]
\begin{subfigure}{.5\textwidth}
\centering
\includegraphics[width=1.0\linewidth]{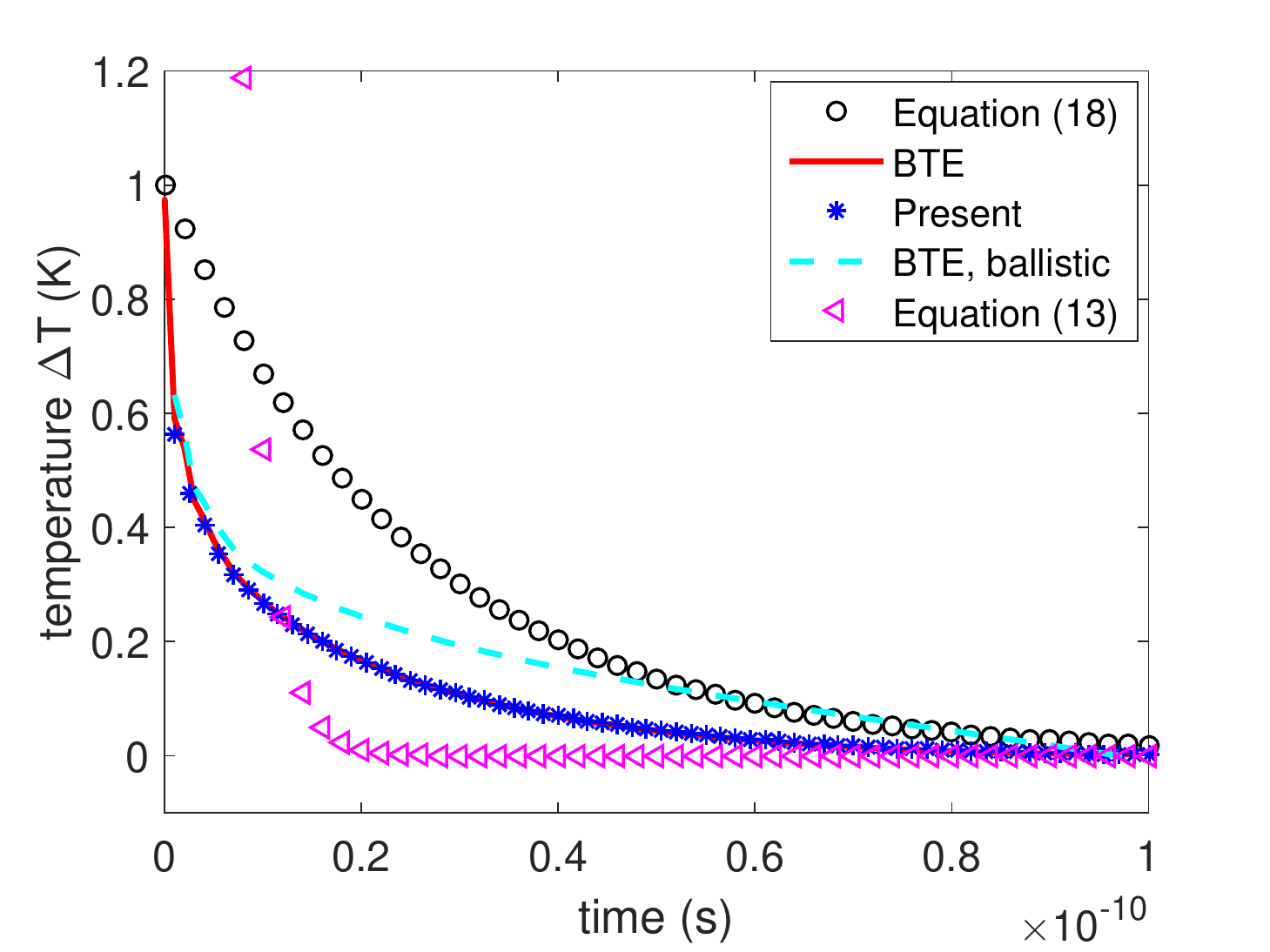}
\caption{Relaxation profile for $L= 10\ nm$ }
\end{subfigure}
\begin{subfigure}{.5\textwidth}
\centering
\includegraphics[width=1.0\linewidth]{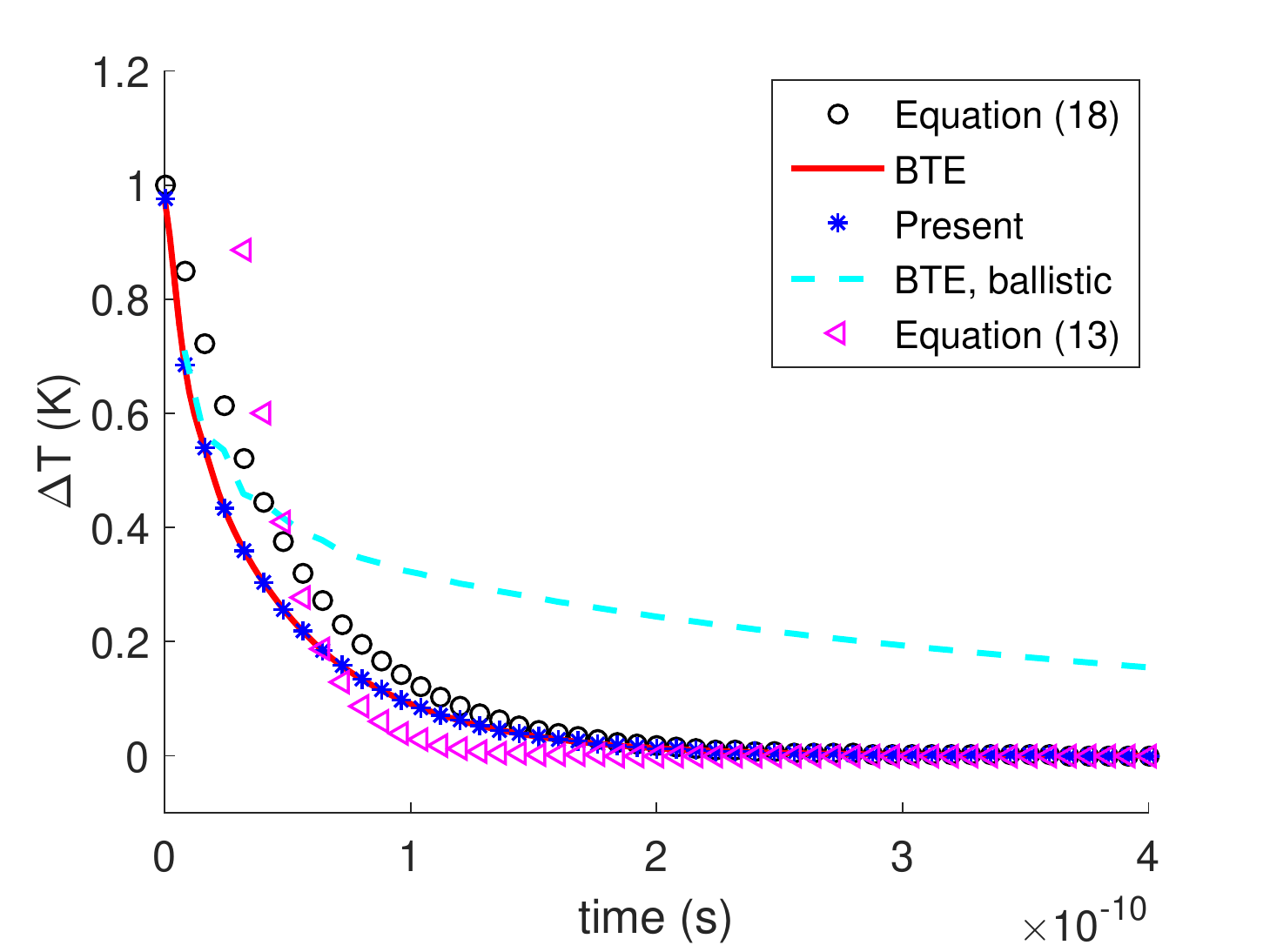}
\caption{Relaxation profile for $L= 100\ nm$ }
\end{subfigure}
\begin{subfigure}{.5\textwidth}
\centering
\includegraphics[width=1.0\linewidth]{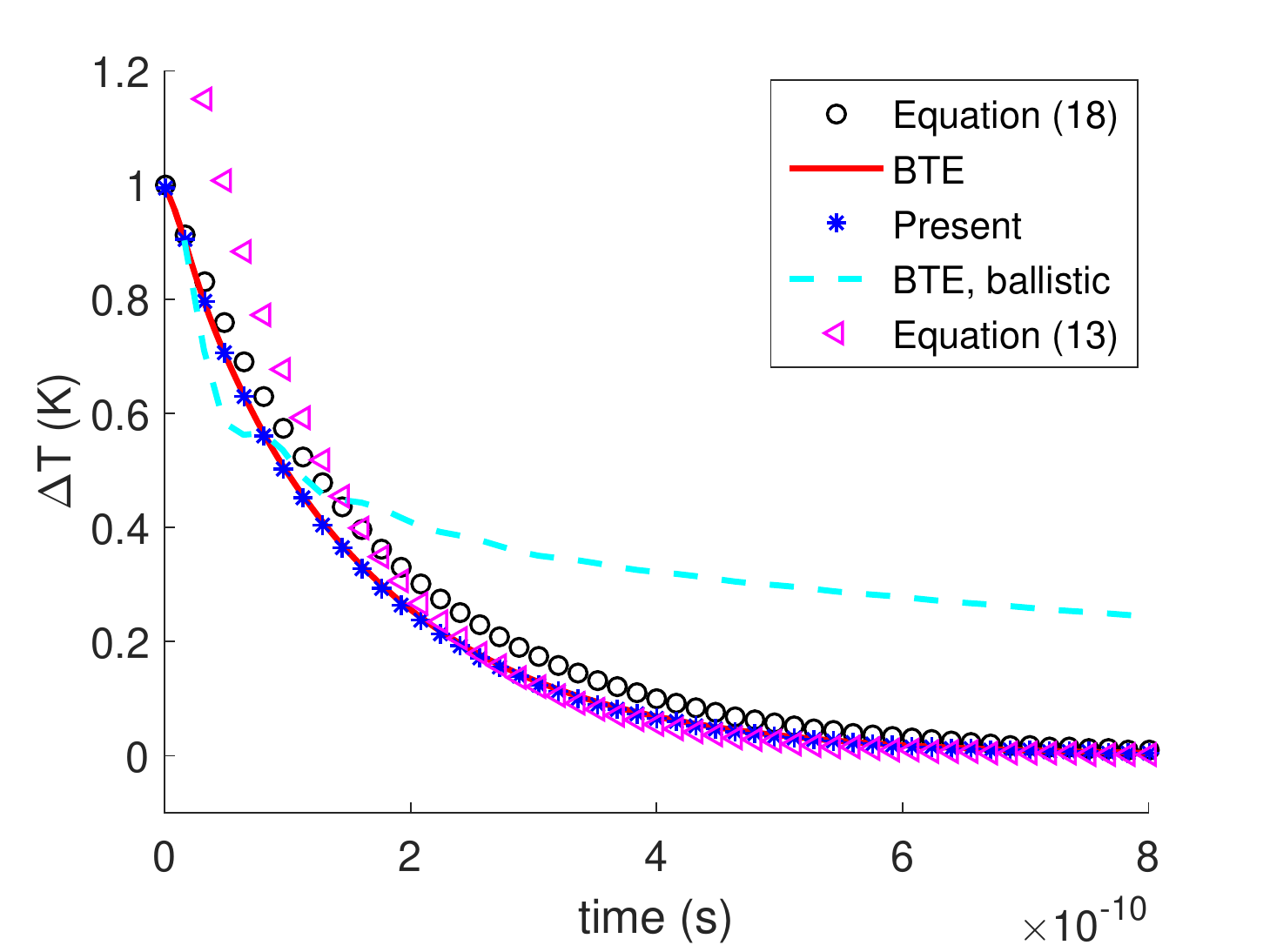}
\caption{Relaxation profile for $L= 400\ nm$ }
\end{subfigure}
\begin{subfigure}{.5\textwidth}
\centering
\includegraphics[width=1.0\linewidth]{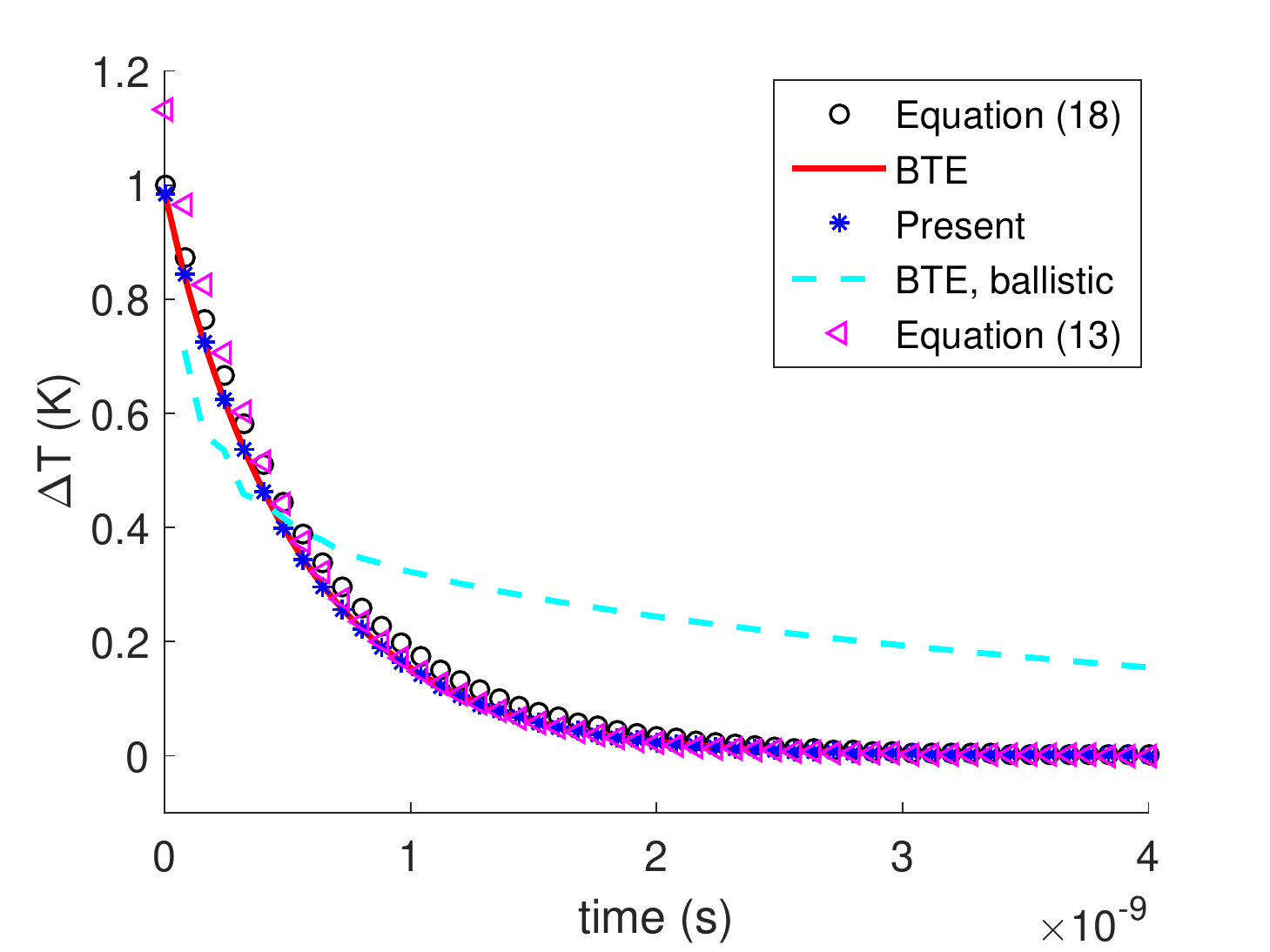}
\caption{Relaxation profile for $L= 1\ \mu m$  }
\end{subfigure}
\caption{Comparison between the method proposed in the present paper and various Fourier-based methods. Temperature relaxation profiles are shown for four grating wavelengths and compared to BTE solution for the Holland material model (denoted ``BTE"). The prediction of the method proposed here, denoted as ``present", is given by solution of the Boltzmann equation based on the reconstructed relaxation time data of figure \ref{compwithanalytical}. Solutions based on equations \eqref{T_quasi} and \eqref{T_quasi_k_eff} are denoted ``Equation \eqref{T_quasi}" and ``Equation \eqref{T_quasi_k_eff}", respectively.}
\label{Temp_comparison}
\end{figure}

The above discussion is validated in figures \ref{Temp_comparison_2TA} and \ref{Temp_comparison}; the figures compare various Fourier-based relaxation profiles to the BTE solution of the grating relaxation problem for various values of the wavelength $L$. In addition to solution \eqref{T_quasi}, we also consider the traditional effective thermal conductivity approach given by \cite{Hua2014},
\begin{equation}
\Delta T= \exp \left( \frac{2 \pi i x}{L} \right) \exp \left(-\left( \frac{2 \pi} {L} \right)^2 \frac{\kappa_{mod}} {C} t \right) .
\label{T_quasi_k_eff}
\end{equation}
The figures also show the solution of the same problem obtained via IFFT of \eqref{Analytical} using the {\it reconstructed} $\tau_{\omega}$ data (as shown in figure \ref{compwithanalytical} for the Holland model and figure \ref{compwithanalytical_2TA} for the ab initio model). Figure \ref{Temp_comparison_2TA} shows results for the ab initio material model ($\langle \Lambda_{\omega} \rangle_{ab}= 95.7\ nm$) and figure \ref{Temp_comparison} for the Holland model ($\langle \Lambda_{\omega} \rangle_{H}= 85.5\ nm$). Here, $\langle \Lambda_{\omega} \rangle$ was calculated using the procedure outlined in \cite{JPThesis}.

In both figures, we observe that the BTE solution based on reconstructed relaxation times, referred to as ``present", is the only solution that is able to predict the correct temperature profiles (IFFT of \eqref{Analytical}, denoted by ``BTE") in all $K\!n$-regimes, for all times. Referring to figure \ref{Temp_comparison_2TA}, at $L= 10\ nm$, the ballistic solution of the BTE (provided in Appendix D) is also able to predict the temperature profile, although for $t \gtrsim 1.5\ ps$ some discrepancies are observable (scattering is no longer completely negligible); Fourier-based approximations are inaccurate. For $L= 100\ nm$ ($K\!n \sim 0.96$), none of the Fourier-based approximations are able to predict the correct temperature profile; although solution \eqref{T_quasi} provides a significantly better approximation compared to expression \eqref{T_quasi_k_eff} at late times, the BTE solution is clearly not of exponential form at early and moderate times (in other words, although \eqref{T_quasi} appears to be close to the BTE solution, this is partly because both go to 0 at late times). At $L= 400\ nm$ ($K\!n \sim 0.24$) the agreement between \eqref{T_quasi} and the BTE solution is good; equation \eqref{T_quasi_k_eff} remains inaccurate. The comparison at $L= 1\ \mu m$  ($K\!n \sim 0.096$) shows good agreement between equation \eqref{T_quasi} and the BTE solution; equation \eqref{T_quasi_k_eff} also becomes a reasonable approximation. Further comparison (not shown here) shows that at $L= 10\ \mu m$ ($K\!n \sim 0.0096$), both Fourier-based solutions are able to reproduce the Boltzmann solution. 

A comparison for the Holland model can be found in figure \ref{Temp_comparison}. The results are similar, but generally reveal worse performance by equation \eqref{T_quasi}  in the critical $K\!n= 0.2-1$ range. Also, the ballistic behavior is observed at smaller wavelengths (approximately $L= 1\ nm$--not shown here).

In summary, although relation \eqref{T_quasi} shows that a regime featuring exponential relaxation is possible, this is only physically realizable (and measurable) for $K\!n^2 \ll 1$. This is clearly more favorable than the typical requirement ($K\!n \ll 1$ for homogeneous materials, $K\!n \lll 1$ in the presence of boundaries \cite{JPasymptotic}) for validity of Fourier's law uncorrected; the improved range of validity is due to the corrections \eqref{k_quasi}-\eqref{C_eff}, which are available, in part, due to the simplicity of the TTG problem and geometry. Although extensions of Fourier approaches are sometimes possible (e.g. see \cite{JPasymptotic} for the role of boundaries), these are not always guaranteed to exist or be tractable and their range of validity cannot be expected to extend to $K\!n \sim 1$, except fortuitously.

With the above considered, for the TTG case, the effective thermal conductivity construction and associated solution \eqref{T_quasi} is only strictly theoretically justified in the limit $K\!n^2 \ll 1$. In some cases this coincides with the range in which experiments operate \cite{Maznev2011}, in which case reconstruction using a diffusive approximation may be acceptable  provided:
\begin{itemize}
\item  the correct asymptotic relation (i.e. equation \eqref{T_quasi}) is used
\item some loss of accuracy outside the range over which experimental data is available is acceptable
\end{itemize}
We also point out that the validation in \cite{Minnich2012} featured wavelengths satisfying $K\!n^2 \ll 1$.
In general, however, the requirements $K\!n^2 \ll 1$ and $t \gg \tau_{\omega}$ are very restrictive, especially when considering that typical pump-probe response time is on the order of a few nanoseconds \cite{Lingping_nature}. In contrast, the method proposed here introduces no lengthscale or timescale restriction; moreover, if input data is available over a limited range of wavelengths, the quality of the reconstruction does not suffer significantly. 

\section{Summary and outlook} 
\label{conclusion}
By analyzing the TTG response in detail we have shown that, as expected, reconstruction using the effective thermal conductivity concept is limited to late (compared to the mean free time) times and large (compared to the mean free path) lengthscales ($K\!n^2 \ll 1$ for the TTG in particular). Using input data that does not satisfy these requirements will lead to reconstruction error; on the other hand, using data in a limited range of characteristic lengthscales also leads to reconstruction error \cite{Minnich2012}. 

In this paper we have  proposed an alternative approach to the reconstruction problem that avoids the limitations associated with the effective thermal conductivity approach. The proposed method assumes knowledge of the phonon group velocities and poses the reconstruction as an optimization problem seeking the function $\tau_{\omega}$ that minimizes the discrepancy between the observed experimental data and the solution of the BTE as applied to the experimental process. Although much research and improvements remain to be made, one of the main contributions of the present paper is to show that reconstruction based on a more rigorous foundation (which does not make use of the assumption of Fourier behavior) is possible.

The proposed formulation is sufficiently general to accept solutions of the BTE obtained by any means. In fact, we envision applications to problems including interfaces between materials \cite{Interface} (and where the reconstruction will include inference of interface properties) as direct extensions of the present work.

The optimization process is achieved by the Nelder-Mead algorithm that does not require gradient calculation and is thus robust to noise. Our results verify that reconstruction is robust both in the presence of noise in the input (experimental) data and the BTE solutions when the latter are obtained by MC simulation. Although MC simulations are considerably more expensive, they are, in fact, the more relevant to applications of practical interest (e.g. \cite{Lingping_nature}) which are in general high-dimensional and not expected to lend themselves to analytical solutions. 
The cost of MC simulations is partially mitigated by their excellent parallel efficiency. Moreover, in cases where the experimental measurement is limited to one (or a few) discrete spatial locations, the MC simulation cost can be made independent of dimensionality using techniques such as adjoint formulations. 

Our validation tests started from multiple distinct initial conditions thus subjecting the optimization process to a worst-case scenario, since in most cases some a-priori information on the nature of the solution is expected to exist. In the present case, the lack of prior information was overcome by starting the optimization process, as a first stage, from a sweep of different initial conditions and choosing the result with the lowest objective function value as the initial condition for a second optimization stage. In the case of MC simulations, this approach was possible due to the robustness of the NM algorithm to noise, which enabled us to perform the first two stages of calculations using very cheap (noisy) simulations. In general, a-priori  information could be very useful in ensuring convergence to the correct solution, which due to the highly non-convex nature of the minimization problem is not guaranteed. Future research will focus on formulations which exploit a-priori information on the nature of the solution as well as alternative  optimization algorithms to improve convergence rates and accuracy. 

Our results also show that the reconstruction process is very robust with respect to the time window over which the material response is observed, in contrast to Fourier-based approaches which assume a late-time response. More specifically, by matching the material response to BTE solutions, the reconstruction is successful with response data limited to very early times (e.g. on the order of 5 nanoseconds) that are typical in pump-probe experiments \cite{Lingping_nature} but do not satisfy assumptions required for diffusive behavior to set in. 

Validation of the proposed methodology using experimental data of thermal relaxation processes will be the subject of a future publication.

\section*{Acknowledgment}
The authors would like to thank V. Chiloyan, S. Huberman, A. Maznev and L. Zeng for many useful comments and discussions. This work was supported by the Solid-State Solar-Thermal Energy Conversion Center (S3TEC), an Energy Frontier Research Center funded by the U.S. Department of Energy, Office of Science, Basic Energy Sciences under Award\# DE-SC0001299 and DE-FG02-09ER46577.

\appendix
\section*{Appendix A: Parameters of piece-wise linear model}
The ranges at which each line or polynomial in relationship \eqref{parametrization_2TA} is active, $X^{S}_{j}$s, are calculated via the following equations
\begin{equation}
X^{S}_{0}= \omega^{S}_{0} , \ \ X^{S}_{2j}= \omega^{S}_{j}+ \Delta , \ \ X^{S}_{2j-1}= \omega^{S}_{j}- \Delta , \ \ X^{S}_{2M-1}= \omega^{S}_{M} ,
\label{optim_parameter_1}
\end{equation}
where $S \in \{LA, TA_1, TA_2\}$, $j \in \{1,..., M-1\}$, and the parameter $\Delta$ is the range of frequency at which each polynomial is active which is taken to be $\Delta= 5 \times 10^{12} \ rad/s$ in this work. The coefficients of polynomials, $a^{S}_{j}$, $b^{S}_{j}$, $c^{S}_{j}$, and $d^{S}_{j}$ are calculated from the following equations
\begin{align}
a^{S}_{j} &= -\frac{\log \left( 1- \left( \frac{\Delta} {\omega^{S}_{j}} \right)^2 \right) } {\left[ \log \left( \frac{\omega^{S}_{j}+ \Delta} {\omega^{S}_{j}- \Delta} \right) \right]^3} \left[ \frac{\log \left( \frac{\tau^{S}_{\omega_{j+1}}}{\tau^{S}_{\omega_{j}}} \right)} { \log \left( \frac{\omega^{S}_{j+1}}{\omega^{S}_{j}} \right) }- \frac{\log \left( \frac{\tau^{S}_{\omega_{j}}} {\tau^{S}_{\omega_{j-1}}} \right)} { \log \left( \frac{\omega^{S}_{j}}{\omega^{S}_{j-1}} \right) } \right] , \nonumber \\
b^{S}_{j} &= 0.5 \left\{ \frac{1} { \log \left( \frac{\omega^{S}_{j}+ \Delta} {\omega^{S}_{j}- \Delta} \right) } \left[ \frac{\log \left( \frac{\tau^{S}_{\omega_{j+1}}} {\tau^{S}_{\omega_{j}}} \right)} { \log \left( \frac{\omega^{S}_{j+1}} {\omega^{S}_{j}} \right) }- \frac{\log \left( \frac{\tau^{S}_{\omega_{j}}} {\tau^{S}_{\omega_{j-1}}} \right)} { \log \left( \frac{\omega^{S}_{j}} {\omega^{S}_{j-1}} \right) } \right]- 3 a^{S}_{j} \log \left( \left( \omega^{S}_{j} \right)^2- \Delta^2 \right) \right\} , \nonumber \\ c^{S}_{j} &= \frac{\log \left( \frac{\tau^{S}_{\omega_{j}}} {\tau^{S}_{\omega_{j-1}}} \right)} { \log \left( \frac{\omega^{S}_{j}} {\omega^{S}_{j-1}} \right) }- 3 a^{S}_{j} \left[ \log \left( \omega^{S}_{j}- \Delta \right) \right]^2- 2 b^{S}_{j} \log \left( \omega^{S}_{j}- \Delta \right) , \nonumber \\ d^{S}_{j} &= \log (\tau^{S}_{\omega_{j-1}})- \log (\omega^{S}_{j-1}) \frac{\log \left( \frac{\tau^{S}_{\omega_{j}}} {\tau^{S}_{\omega_{j-1}}} \right)} { \log \left( \frac{\omega^{S}_{j}} {\omega^{S}_{j-1}} \right) } + \nonumber \\ & \hspace{80pt} 0.5 \frac{ \left[ \log \left( \omega^{S}_{j}- \Delta \right) \right]^2 } { \log \left( \frac{\omega^{S}_{j}+ \Delta} {\omega^{S}_{j}- \Delta} \right) } \left[ \frac{\log \left( \frac{\tau^{S}_{\omega_{j+1}}} {\tau^{S}_{\omega_{j}}} \right)} { \log \left( \frac{\omega^{S}_{j+1}} {\omega^{S}_{j}} \right) }- \frac{\log \left( \frac{\tau^{S}_{\omega_{j}}} {\tau^{S}_{\omega_{j-1}}} \right)} { \log \left( \frac{ \omega^{S}_{j}}{\omega^{S}_{j-1}} \right) } \right]- \nonumber \\ & \hspace{80pt} 0.5 a^{S}_{j} \left[ \log \left( \omega^{S}_{j}- \Delta \right) \right]^2 \left[ 3 \log ( \omega^{S}_{j}+ \Delta)- \log (\omega^{S}_{j}- \Delta) \right] ,
\label{optim_parameter_2}
\end{align}
where $j \in \{1,..., M-1\}$ and $S \in \{LA, TA_1, TA_2 \}$.

\section*{Appendix B: Semi-analytical solution for the transient thermal grating}
Here, we derive the semi-analytical frequency-domain solution \eqref{Analytical} used in section \ref{TTG} using Fourier transform in the frequency domain (with respect to the time variable) and wavenumber domain (with respect to the space variable). We will use the subscript $\zeta$ to denote the former and the subscript $\eta$ to denote the latter. By applying a Fourier transform to \eqref{BTE_TTG} in both variables we obtain
\begin{equation} \label{BTE_Fourier}
i \zeta {e}^d_{\zeta \eta}+ i \eta v_{\omega} \cos(\theta) {e}^d_{\zeta \eta}= -\frac{1} {\tau_{\omega}} {e}^d_{\zeta \eta}+ \frac{1}{\tau_{\omega}} (de^{eq}/dT)_{T_{eq}} \Delta \widetilde{T}_{\zeta \eta}+ 2 \pi (de^{eq}/dT)_{T_{eq}} \delta \left( \eta- 2 \pi L^{-1} \right) ,
\end{equation}
which provides the following relationship for ${e}^d_{\zeta \eta}$
\begin{equation} \label{Energy_Fourier}
{e}^d_{\zeta \eta}= \frac{(de^{eq}/dT)_{T_{eq}} \left[\tau_{\omega}^{-1} \Delta \widetilde{T}_{\zeta \eta}+ 2 \pi \delta \left( \eta- 2 \pi L^{-1} \right) \right] } {i \zeta+ i \eta v_{\omega} \cos(\theta)+ \tau_{\omega}^{-1} } .
\end{equation}
Applying Fourier transform in both variables to equation \eqref{pseudotemperature} and replacing ${e}^d_{\zeta \eta}$ with the above result, we obtain the following expression for $\Delta \widetilde{T}_{\zeta \eta}$
\begin{equation}
\int_{\mathbf{\Omega}} \int_{\omega} \left[ \frac{C_\omega} {\tau_{\omega} } \Delta \widetilde{T}_{\zeta \eta} -\frac{(de^{eq}/dT)_{T_{eq}} \left[ \tau_{\omega}^{-1} \Delta \widetilde{T}_{\zeta \eta}+ 2 \pi \delta \left( \eta- 2 \pi L^{-1} \right) \right] } { i \tau_{\omega} \zeta+ i \eta v_{\omega} \tau_{\omega} \cos(\theta) + 1 } D_{\omega} \right] d \omega d \mathbf{\Omega}= 0 , \label{pseudo_Fourier}
\end{equation}
which after integrating over $\theta \in (0, \pi)$ and $\phi \in (0, 2\pi)$ and algebraic simplification leads to the following relationship 
\begin{equation}
\Delta \widetilde{T}_{\zeta \eta}= \frac{\frac{ \pi \delta \left(\eta- 2 \pi L^{-1} \right)} {\eta} \int_{\omega} \frac{i C_\omega } {v_{\omega} \tau_{\omega}} \ln \left( \frac{\tau_{\omega} \zeta- v_{\omega} \tau_{\omega} \eta- i} {\tau_{\omega} \zeta+ v_{\omega} \tau_{\omega} \eta- i} \right) d \omega} { \int_{\omega} \frac{C_\omega} {\tau_{\omega}} d \omega- \frac{1} {2 \eta} \int_{\omega} \frac{i C_\omega } {v_{\omega} \tau_{\omega}^2} \ln \left( \frac{\tau_{\omega} \zeta- v_{\omega} \tau_{\omega} \eta -i} {\tau_{\omega} \zeta+ v_{\omega} \tau_{\omega} \eta- i} \right) d \omega } . \label{pseudo_Fourier2}
\end{equation}
Equation \eqref{pseudo_Fourier2} can be further simplified by applying the inverse Fourier transform in wavenumber domain $\eta$. Noting that equation \eqref{pseudo_Fourier2} is in the form of $\Delta \widetilde{T}_{\zeta \eta}= \delta (\eta- 2 \pi L^{-1} ) G (\eta, \zeta)$, we use the identity
\begin{equation} \label{FourierIntegral}
\frac{1} {2 \pi} \int_{-\infty}^{\infty} \delta \left(\eta- 2 \pi L^{-1} \right) G (\eta, \zeta) \mathrm{e}^{i \eta x} d \eta= \frac{1} {2 \pi} G \left( 2 \pi L^{-1}, \zeta \right) \mathrm{e}^{2 \pi i x/L} ,
\end{equation}
to obtain
\begin{equation}
\Delta \widetilde{T}_{\zeta}= \frac{\mathrm{e}^{2 \pi i x/L} \int_{\omega} S_{\omega} \tau_{\omega} d \omega} {\int_{\omega} \left[ C_{\omega} \tau_{\omega}^{-1}- S_{\omega} \right] d \omega} , \label{pseudo_Final}
\end{equation}
where $S_{\omega}$ is given in \eqref{S_omega}. Equation \eqref{pseudo_Final} is similar to the result obtained in \cite{Hua2014}. 

To obtain an expression for the temperature, we insert \eqref{pseudo_Fourier2} into \eqref{Energy_Fourier} and apply the inverse Fourier transform in wavenumber domain using the identity \eqref{FourierIntegral} to obtain
\begin{equation} \label{Energy_Fourier_half}
e^d_{\zeta}= \frac{ \mathrm{e}^{ 2 \pi i x/L} (de^{eq}/dT)_{T_{eq}}} {i \tau_{\omega} \zeta+ i v_{\omega} \tau_{\omega} \cos(\theta) 2 \pi L^{-1}+ 1} \left[\frac{ \int_{\omega} S_{\omega} \tau_{\omega} d \omega } { \int_{\omega} C_{\omega} \tau_{\omega}^{-1} d \omega- \int_{\omega} S_{\omega} d \omega }+ 1 \right] .
\end{equation}
This relation is then substituted in 
\begin{equation}
\int_{\mathbf{\Omega}} \int_{\omega} \left[ C_{\omega} \Delta {T}_\zeta- {e}^d_\zeta D_{\omega} \right] d \omega d \mathbf{\Omega}= 0 , \label{temperature_Fourier}
\end{equation}
obtained by applying Fourier transform in the frequency domain to \eqref{temperature}. Solving the resulting equation for $\Delta T_{\zeta}$  leads to the expression provided in \eqref{Analytical}.

\section*{Appendix C: Solution of the transient thermal grating for late times}
Introducing the definition $K\!n_\omega= v_{\omega} \tau_{\omega} 2 \pi L^{-1}$ in equation \eqref{S_omega}, provides the following relationship for $S_{\omega}$
\begin{equation}
S_{\omega}= \frac{ i C_{\omega}} {2 K\!n_{\omega} \tau_{\omega} } \ln \left( \frac{\tau_{\omega} \zeta- K\!n_{\omega}- i} {\tau_{\omega} \zeta+ K\!n_{\omega}- i} \right) . \label{S_omega_new}
\end{equation} 
Using a Taylor expansion for $\tau_{\omega} \zeta \ll 1$ of the logarithmic term in \eqref{S_omega_new} we obtain  
\begin{equation}
S_{\omega}= \frac{ i C_{\omega}} {2 K\!n_{\omega} \tau_{\omega} } \left[ -2 i \tan^{-1} \left( K\!n_{\omega} \right)- \frac{2 K\!n_{\omega} \tau_{\omega}} {K\!n_{\omega}^2+ 1 } \zeta+ O \left( \tau_{\omega}^2 \zeta^2 \right) \right] , \label{S_omega_new_Taylor}
\end{equation}
which is the same expression provided in equation \eqref{S_omega_quasi}. If we substitute $S_{\omega}$ in \eqref{Analytical} with \eqref{S_omega_new_Taylor} and neglect terms of $O \left( \tau_{\omega}^2 \zeta^2 \right) $ we obtain
\begin{multline}
\Delta T_{\zeta}= \frac{ \mathrm{e}^{ 2 \pi i x/L}} {C} \Bigg\{ \int_{\omega} C_{\omega} \tau_{\omega} \left[ \frac{ \tan^{-1} \left( K\!n_{\omega} \right) } {K\!n_{\omega}}- \frac{ \tau_{\omega} i \zeta } {K\!n_{\omega}^2+ 1 } \right] d \omega+ \\  \frac{ \left[ \int_{\omega} C_{\omega} \left( \frac{\tan^{-1} \left( K\!n_{\omega} \right)} {K\!n_{\omega} }- \frac{\tau_{\omega} i \zeta} { K\!n_{\omega}^2+ 1 } \right) d \omega \right]^2} { \int_{\omega} \frac{C_{\omega}} {\tau_{\omega} } \left[ 1- \frac{\tan^{-1} \left( K\!n_{\omega} \right)} {K\!n_{\omega}}+ \frac{\tau_{\omega} i \zeta} { K\!n_{\omega}^2+ 1 } \right] d \omega } \Bigg\} .
\label{Analytical_quasi}
\end{multline}
Equation \eqref{Analytical_quasi} is in the general form of
\begin{equation}
\Delta T_{\zeta}= k_1- k_2 i \zeta+ \frac{(k_3- k_4 i \zeta)^2} {k_5+ k_6 i \zeta} ,
\label{Analytical_quasi_2}
\end{equation}
where $k_1$, $k_2$,..., $k_6$ are independent of Fourier transform variable $\zeta$ and depend only on the geometry and the material properties. Equation \eqref{Analytical_quasi_2} can be further simplified in the following form
\begin{equation}
\Delta T_{\zeta}= k_1- \frac{k_4^2 k_5} {k_6^2}- \frac{2 k_3 k_4} {k_6}+ \left( \frac{k_4^2} {k_6}- k_2 \right) i \zeta+ \frac{ \left( k_4 k_5+ k_3 k_6 \right)^2} {k_6^2 (k_5+ k_6 i \zeta)} .
\label{Analytical_quasi_3}
\end{equation}
Applying inverse Fourier transform to \eqref{Analytical_quasi_3} leads to the following result
\begin{equation}
\Delta T= \left( k_1- \frac{ k_4^2 k_5} {k_6^2}- \frac{2 k_3 k_4} {k_6} \right) \delta(t)+ \left( \frac{k_4^2} {k_6}- k_2 \right) \delta'(t)+ \frac{\left( k_4 k_5+ k_3 k_6 \right)^2} {k_6^3} u(t) \mathrm{e}^{-\frac{k_5} {k_6} t} ,
\label{Analytical_quasi_4}
\end{equation}
where $\delta(t)$ and $\delta'(t)$ are the delta Dirac function and its first derivative, respectively, and $u(t)$ is the Heaviside function. At late times (more precisely, $t>0$), only the last term of \eqref{Analytical_quasi_4} is nonzero. Substituting constants $k_1$ through $k_6$ in \eqref{Analytical_quasi_4} with their equivalents from equation \eqref{Analytical_quasi} leads to the result provided in \eqref{T_quasi}.
 
\section*{Appendix D: Analytical solution for transient thermal grating in ballistic limit}
The temperature relaxation profile in the ballistic limit can be obtained by setting $\tau_{\omega} \rightarrow \infty$ in equation \eqref{Analytical} and \eqref{S_omega}. Using this substitution, for $S_{\omega}$ we have
\begin{equation}
S_{\omega}= \frac{i C_\omega L} {4 \pi v_{\omega} \tau_{\omega}^2} \ln \left( \frac{\zeta- v_{\omega} 2 \pi L^{-1}} {\zeta+ v_{\omega} 2 \pi L^{-1}} \right) . \label{S_omega_ballistic}
\end{equation}
Substituting this expression in \eqref{Analytical}, we observe that the second term of \eqref{Analytical} vanishes in this limit (the numerator is $O (\tau_{\omega}^{-2})$, while the denominator is $O (\tau_{\omega}^{-1})$). Therefore, we have
\begin{equation}
\Delta {T}_\zeta= \frac{i L \mathrm{e}^{2 \pi i x/L}} {4 \pi C} \int_{\omega} \frac{C_{\omega} } {v_{\omega} } \ln \left( \frac{\zeta- v_{\omega} 2 \pi L^{-1}} {\zeta+ v_{\omega} 2 \pi L^{-1}} \right) d \omega . \label{Analytical_ballistic}
\end{equation}
The inverse Fourier transform of \eqref{Analytical_ballistic} can be calculated analytically \cite{integral_table} and gives
\begin{equation}
\Delta T= \frac{L \mathrm{e}^{2 \pi i x/L}} {2 \pi C t} \int_{\omega} \frac{C_\omega } {v_{\omega} } \sin \left( v_{\omega} 2 \pi L^{-1} t \right) d \omega .
\label{Analytical_ballistic_time}
\end{equation}
\end{document}